\documentclass[useAMS,usenatbib]{mn2e}
\usepackage{graphicx}
\usepackage{multicol}
\usepackage{enumerate}
\usepackage{amsmath}
\usepackage{pdflscape}
\usepackage{longtable}
\usepackage[section]{placeins}

\usepackage{journal_names}

\title[Optical counterparts to GBS X-ray sources] {The \textit{Chandra Galactic Bulge Survey}: optical catalogue and point-source counterparts to X-ray sources}
\author[Wevers et al.]{T. Wevers$^{1}$\thanks{Email: t.wevers@astro.ru.nl}, S. T. Hodgkin$^{2}$, P. G. Jonker$^{3,1}$, C. Bassa$^{4}$, G. Nelemans$^{1,5}$, \newauthor T. van Grunsven$^{3}$, E. A. Gonzalez-Solares$^{2}$, M. A. P. Torres$^{3}$, C. Heinke$^{6}$, \newauthor D. Steeghs$^{7}$, T. J. Maccarone$^{8}$,  C. Britt$^{9,8}$, R. I. Hynes$^{9}$, C. Johnson$^{9}$,  Jianfeng Wu$^{10,11}$ \\\\
$^{1}$Department of Astrophysics/IMAPP, Radboud University Nijmegen, P.O. box 9010, 6500 GL Nijmegen, The Netherlands\\
$^{2}$Institute of Astronomy, Madingley Road, Cambridge CB3 0HA, UK\\
$^{3}$SRON, Netherlands Institute for Space Research, Sorbonnelaan 2, 3584 CA Utrecht, the Netherlands\\
$^{4}$ASTRON, the Netherlands Institute for Radio Astronomy, Postbus 2, 7990 AA Dwingeloo, The Netherlands\\
$^{5}$Institute for Astronomy, KU Leuven, Celestijnenlaan 200D, 3001 Leuven, Belgium\\
$^{6}$Dept. of Physics, University of Alberta, CCLS 4-183, Edmonton, AB T6G 2E1, Canada\\
$^{7}$Department of Physics, University of Warwick, Coventry CV4 7AL, UK\\
$^{8}$Department of Physics, Texas Tech University, box 41051, Lubbock, TX 79409-1051, USA\\
$^{9}$Department of Physics and Astronomy, Louisiana State University, Baton Rouge, LA 70803-4001, USA\\
$^{10}$Harvard-Smithsonian Center for Astrophysics, 60 Garden Street, Cambridge, MA 02138, USA\\
$^{11}$Department of Astronomy, University of Michigan, 1085 S University Ave, Ann Arbor, MI, 48109, USA
}

\begin{document}
\date{Accepted 2016 March 12. Received 2016 March 1; in original form 2016 January 29}
\pagerange{\pageref{firstpage}--\pageref{lastpage}} \pubyear{2016}
\maketitle
\label{firstpage}

\begin{abstract}
As part of the Chandra Galactic Bulge Survey (GBS), we present a catalogue of optical sources in the GBS footprint. This consists of two regions centered at Galactic latitude b = 1.5$^{\circ}$ above and below the Galactic Centre, spanning (l $\times$ b) = (6$^{\circ}$$\times$1$^{\circ}$). The catalogue consists of 2 or more epochs of observations for each line of sight in r$^{\prime}$, i$^{\prime}$ and H$\alpha$ filters. The catalogue is complete down to r$^{\prime}$ = 20.2 and i$^{\prime}$ = 19.2 mag; the mean 5$\sigma$ depth is r$^{\prime}$ = 22.5 and i$^{\prime}$ = 21.1 mag. The mean root-mean-square residuals of the astrometric solutions is 0.04 arcsec. We cross-correlate this optical catalogue with the 1640 unique X-ray sources detected in Chandra observations of the GBS area, and find candidate optical counterparts to 1480 X-ray sources. We use a false alarm probability analysis to estimate the contamination by interlopers, and expect $\sim$ 10 per cent of optical counterparts to be chance alignments. To determine the most likely counterpart for each X-ray source, we compute the likelihood ratio for all optical sources within the 4$\sigma$ X-ray error circle. This analysis yields 1480 potential counterparts ($\sim$ 90 per cent of the sample). 584 counterparts have saturated photometry (r$^{\prime} \leq 17$, i$^{\prime} \leq 16$), indicating these objects are likely foreground sources and the real counterparts. 171 candidate counterparts are detected only in the i$^{\prime}$-band. These sources are good qLMXB and CV candidates as they are X-ray bright and likely located in the Bulge.

\end{abstract}
\begin{keywords}
catalogues - optical: general - surveys - X-rays: binaries - Galaxy: bulge
\end{keywords}


\section{Introduction}
\label{sec:introduction}
The Chandra Galactic Bulge Survey (GBS; \citeauthor{Jonker2011} \citeyear{Jonker2011}, \citeyear{Jonker2014}) is a multi-wavelength survey (including X-ray, near infrared (NIR) and optical wavelengths) of the Galactic Bulge. The survey area spans two regions of 6$^{\circ}$$\times$1$^{\circ}$ above and below the Galactic plane. The Northern strip runs from Galactic longitude $-3^{\circ} \leq\ l\ \leq\ 3^{\circ}$ and Galactic latitude $1^{\circ} \leq\ b \leq\ 2^{\circ}$, while the Southern strip has $-1^{\circ} \leq\ b \leq\ -2^{\circ}$.
The GBS was designed to detect X-ray sources and their optical/NIR counterparts, to allow the classification of discovered X-ray sources based on multi-wavelength photometric and spectroscopic observations. The main advantage of the Bulge (over the Galactic Centre) is that dust extinction decreases quickly as one moves out of the plane, significantly increasing the fraction of X-ray sources for which counterparts at other wavelengths can be identified. At the same time, the lower source densities reduce the negative effects of source confusion and crowding. The GBS X-ray source catalogue consists of 1640 unique sources \citep{Jonker2014} for which more than 3 X-ray photons were detected in 2 ks exposures in the 0.3 - 7 keV channel of the Chandra X-ray observatory. The flux limit of the X-ray observations is $(1 - 3) \times 10^{-14}$ erg cm$^{-2}$ s$^{-1}$.

The GBS was designed with 2 main science goals in mind, both of which require a substantial number of X-ray sources to realise (\citeauthor{Jonker2011} \citeyear{Jonker2011}, \citeyear{Jonker2014}). The first science goal is the (model-independent) measurement of neutron star (NS) and black hole (BH) masses to constrain the NS equation of state and BH formation channels. Model-independent mass measurements can only be made in eclipsing X-ray binaries, so we need to identify systems that are suitable for dynamical studies. Given the required high inclination angle, these systems are rare. The GBS is expected to uncover $\sim$ 120-200 LMXBs, so we expect to discover at least a few eclipsing systems.

The second goal of the GBS is to address X-ray binary formation and evolution, in particular the common envelope (CE) evolution of binary stars. The efficiency of CE interactions can be inferred by for example deriving the number ratio of low-mass X-ray binaries (LMXBs) to cataclysmic variables (CVs) (see e.g. \citeauthor{Iben1993} \citeyear{Iben1993}). Studying formation and evolution channels requires comparing the source populations of the detected X-ray sources with binary population synthesis models (see e.g. \citeauthor{vanHaaften2015} \citeyear{vanHaaften2015}), and a large and homogeneously selected sample is critical in this respect. Both of these science goals depend on the identification and classification of the correct multi-wavelength counterpart.

The depth of the Chandra X-ray observations was chosen to optimise the discovery of quiescent LMXBs (qLMXBs) with respect to CVs. At present, the population of known LMXBs consists of persistent and transient systems, with almost no systems discovered in quiescence before entering an outburst. Obtaining a well-defined sample which does not suffer from observational biases is crucial for understanding their formation scenarios. \citet{Jonker2011} estimated that there are 120-200 qLMXBs in the GBS area that have an optical/NIR counterpart bright enough to be discovered in the complementary optical/NIR surveys.

There is an ongoing effort of multi-wavelength and variability studies to identify and characterise the counterparts of the X-ray sources discovered in the GBS. Previous photometric studies have looked at the brightest optical counterparts among Tycho-2 stars (brighter than V=12 mag, \citeauthor{Hynes2012} \citeyear{Hynes2012}) and among the variable stars found in the Optical Gravitational Lensing Experiment \citep{Udalski2012}. All but a handful of the objects identified by these authors are brighter than I=17 mag. The work presented here significantly extends the search for optical counterparts down to limiting magnitudes of r$^{\prime} \leq$ 22.5, i$^{\prime} \leq$ 21.1 mag (mean 5$\sigma$ detection limits) by using observations taken with the Mosaic-II camera mounted at the 4-m Victor M. Blanco telescope at Cerro Tololo Inter-American Observatory (CTIO), Chile. A complimentary search for photometric variables has been performed in the r$^{\prime}$-band with the Mosaic-II camera, but with a shallower depth \citep{Britt2014}.

In addition, attention has also been given to bright radio counterparts \citep{Maccarone2012} and NIR counterparts \citep{Greiss2014}. Some particularly interesting objects have been identified. For example, \citet{Ratti2013} found a new long orbital period CV in a low accretion state, and \citet{Hynes2014} identified a carbon star likely associated with a symbiotic binary. \citet{Torres2014} have found a large sample of accreting binaries using medium-resolution spectroscopic follow-up from H$\alpha$ emission-line selected GBS sources. \citet{Wu2015} found a number of accreting binaries that do not obviously exhibit the characteristic H$\alpha$ emission lines in their spectra. Only after the optimal subtraction of the companion star spectrum can the H$\alpha$ emission line be observed. \\
In this work, we will use deep optical observations of the Galactic Bulge to generate an optical source catalogue which we cross-correlate with the existing GBS X-ray catalogue. 

In Section \ref{sec:opticalcatalog} we present the optical observations and explain how we generate the optical catalogue. In Section \ref{sec:xmatch} we cross-correlate this optical catalogue with the X-ray observations, and we identify the most likely counterparts. The results are presented in Section \ref{sec:results}, and we explore the population properties of the most likely counterparts. We summarise our findings in Section \ref{sec:summary}.


\section{Catalogue of optical sources in the GBS area}
\label{sec:opticalcatalog}
\subsection{Observations}
The optical observations were taken during 10 nights, from 2006 June 20 to 30, using the MOSAIC-II imager mounted in the prime focus of the 4-m Victor M. Blanco telescope at CTIO, Chile. The MOSAIC-II instrument consists of a mosaic of 8 CCDs with a total field of view of 36$\times$36 arcmin and has a pixel-scale of 0.27 arcsec/pixel.  

\begin{figure*}
\includegraphics[width=20cm, height=10cm]{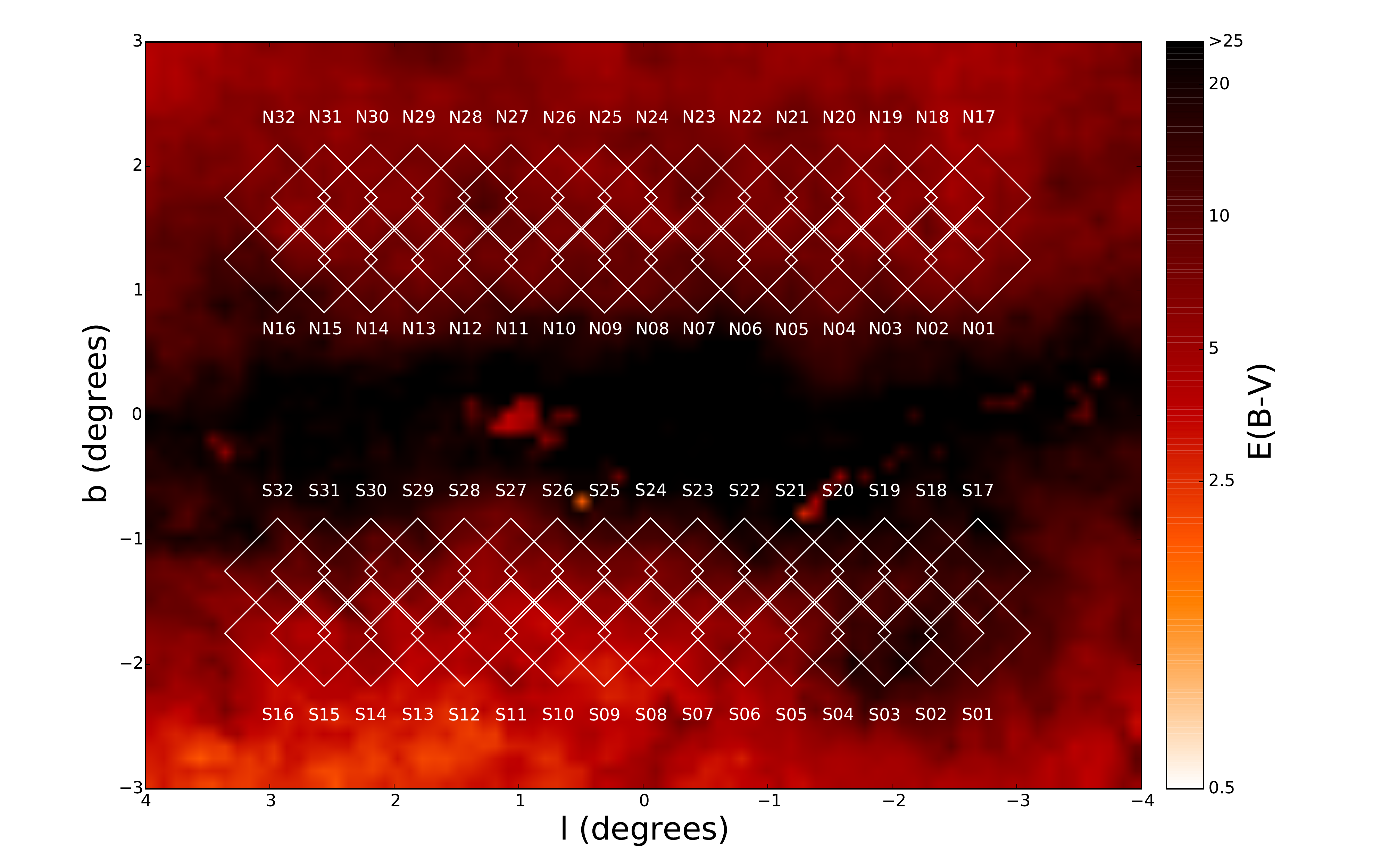}
\caption{Dust map of an 8 by 6 degree region around the Galactic Centre, showing the fields that were observed in the optical as part of the GBS (white squares). The total area covered is approximately 12 deg$^2$. Each rectangle represents one field, which has been observed twice (once with a small offset to cover most of the gaps between the CCDs, not shown on the figure). Two or more fields partially overlap in all lines of sight. The colour-scale traces the interstellar extinction integrated to a distance of 8 kpc (taken from \citet{Schultheis2014}).}
\label{fig:opticalcoverage}
\end{figure*}

\begin{figure}
\includegraphics[width=9cm, keepaspectratio]{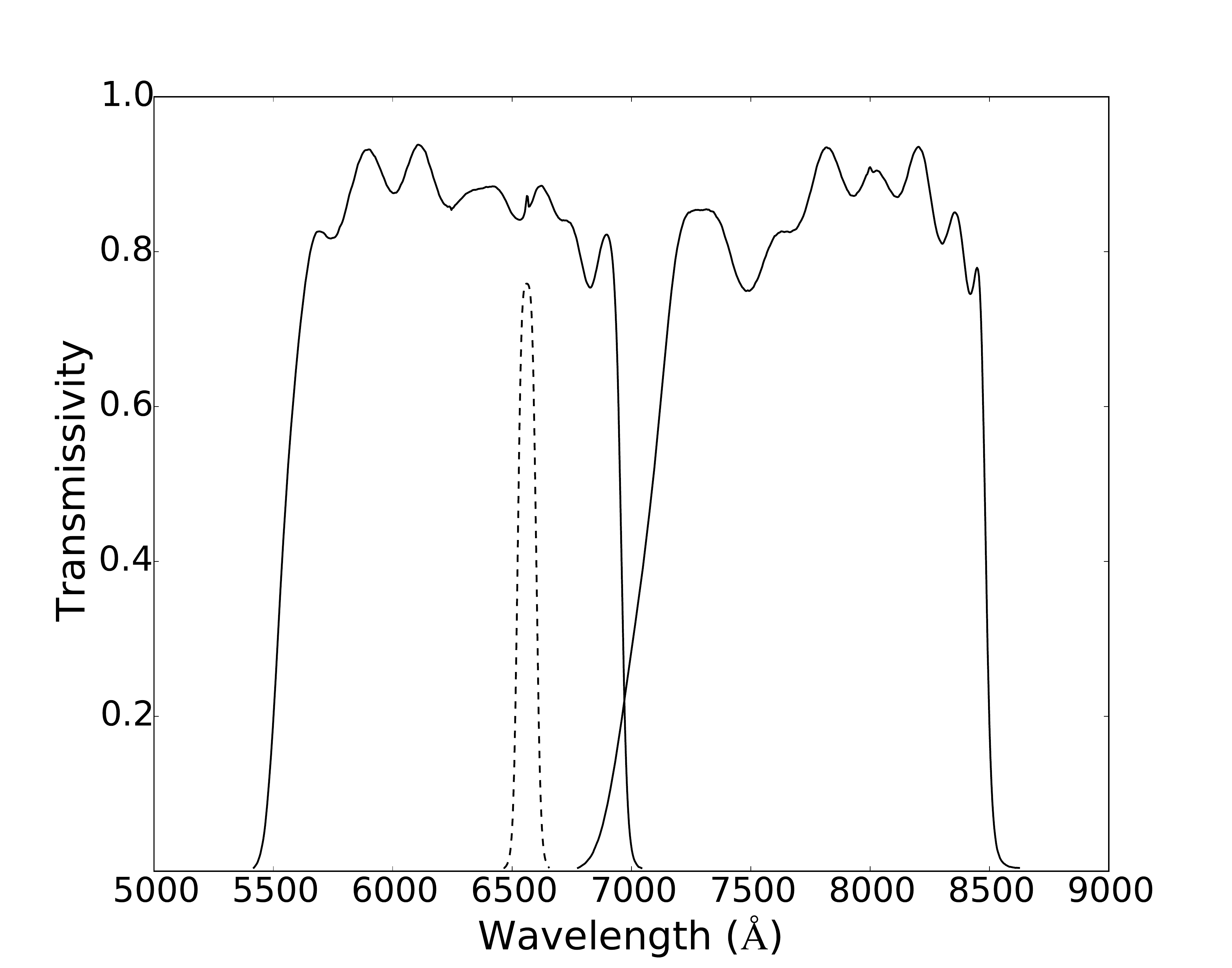}
\caption[captioncaptioncaption]{Transmission profiles of the r$^{\prime}$ (left solid curve), i$^{\prime}$ (right solid curve) and H$\alpha$ (dashed) CTIO filters used in this study.}
\label{fig:transmissionprofiles}
\end{figure}

The 12 deg$^2$ of the GBS area were covered in 64 pointings. Figure \ref{fig:opticalcoverage} shows the layout of the optical survey overlaid on an extinction map by \citet{Schultheis2014}, integrated to a distance of 8 kpc (roughly the distance of the Galactic Centre, \citeauthor{Reid2014} \citeyear{Reid2014}). This illustrates how quickly the interstellar reddening decreases as one moves out of the Galactic plane, and how this survey evades the most reddened lines of sight. 
Each pointing consists of 2 sets of observations, where the second set is offset by 1.2 arcminutes in both right ascension and declination with respect to the first, to almost fully cover the gaps between the CCDs. Each field is observed in 3 filters: r$^{\prime}$, i$^{\prime}$ and H$\alpha$. The filter transmission profiles\footnote{http://svo2.cab.inta-csic.es/svo/theory/fps3/index.php?mode\\=browse$\&$gname=CTIO} are shown in Figure \ref{fig:transmissionprofiles}. To perform the photometric calibration, we have observed standard star fields containing Landolt stars \citep{Landolt1992}.

The offset observation in a given filter is taken right after the first one, but offset in right ascension and declination by $\sim$ 1.2 arcmin. The exposure times are 120, 180 and 480 seconds for r$^{\prime}$, i$^{\prime}$ and H$\alpha$, respectively. Additionally one short 10 second r$^{\prime}$-band exposure was taken for the astrometric calibration. Table \ref{tbl:pointingcentres} lists the pointing centres of the observations, together with the seeing at the time the data were taken. Some fields were observed twice because of bad observing conditions. In that case we use the observations with the best observed seeing, which varied between 0.7 and 1.9 arcsec with a median of 1.06 arcsec.

\begin{table}
 \centering
  \caption{List of the pointing centre (in Galactic coordinates) and the seeing towards each field in the r$^{\prime}$-band. The median seeing during the observing run is 1.06 arcsec. The observing date is the modified Julian date (MJD) of the r$^{\prime}$-band observation, which is the middle exposure of the (i$^{\prime}$, r$^{\prime}$, H$\alpha$) observing sequence for each field.}
  \begin{tabular}{ccccc}
  \hline
  Field & l  ($^{\circ}$) & b ($^{\circ}$) & Seeing ($^{\prime\prime}$) & Date (MJD in days) \\
  \hline\hline
S01 & -2.81051 & -1.74992 & 1.44&53907.0205 \\ 
S02 & -2.43564 & -1.75001 & 1.34&53907.0543 \\ 
S03 & -2.06136 & -1.75097 & 1.26&53907.0813 \\ 
S04 & -1.68859 & -1.74980 & 1.16&53907.1183 \\ 
S05 & -1.31301 & -1.75081 & 1.88&53907.1462 \\ 
S06 & -0.93649 & -1.74999 & 1.19&53908.0257 \\ 
S07 & -0.56141 & -1.75049 & 1.17&53908.0557 \\ 
S08 & -0.18711 & -1.75096 & 1.21&53908.0945 \\ 
S09 & 0.18876 & -1.75271 & 1.12&53908.1211 \\ 
S10 & 0.56157 & -1.75020 & 0.82&53908.1528 \\ 
S11 & 0.93734 & -1.75052 & 0.81&53908.1806 \\ 
S12 & 1.31311 & -1.75129 & 0.95&53908.2084 \\ 
S13 & 1.68817 & -1.74995 & 0.79&53908.2353 \\ 
S14 & 2.06341 & -1.75031 & 0.84&53908.2624 \\ 
S15 & 2.43978 & -1.75109 & 1.05&53908.2898 \\ 
S16 & 2.81217 & -1.75022 & 1.01&53908.3185 \\ 
S17 & -2.81219 & -1.25085 & 1.16&53907.1908 \\ 
S18 & -2.43631 & -1.24988 & 1.09&53907.2184 \\ 
S19 & -2.06201 & -1.24928 & 1.03&53907.2466 \\ 
S20 & -1.68790 & -1.25038 & 1.16&53915.1378 \\ 
S21 & -1.31129 & -1.25102 & 1.18&53907.3022 \\ 
S22 & -0.93712 & -1.25117 & 1.20&53907.3304 \\ 
S23 & -0.56147 & -1.25034 & 1.18&53915.2814 \\ 
S24 & -0.18692 & -1.25044 & 0.96&53915.2481 \\ 
S25 & 0.18867 & -1.25051 & 1.48&53911.2628 \\ 
S26 & 0.56312 & -1.25110 & 1.68&53911.2909 \\ 
S27 & 0.93912 & -1.25033 & 1.06&53912.0313 \\ 
S28 & 1.31323 & -1.25076 & 0.87&53912.0655 \\ 
S29 & 1.68674 & -1.25029 & 0.88&53912.0929 \\ 
S30 & 2.06290 & -1.25089 & 1.04&53912.1195 \\ 
S31 & 2.43777 & -1.25137 & 1.08&53912.1463 \\ 
S32 & 2.81246 & -1.25062 & 1.05&53912.1729 \\ 
N01 & -2.81103 & 1.24974 & 0.92&53912.2039 \\ 
N02 & -2.43660 & 1.24999 & 1.54&53912.2326 \\ 
N03 & -2.05993 & 1.24929 & 1.23&53912.2872 \\ 
N04 & -1.68874 & 1.24698 & 1.16&53912.3142 \\ 
N05 & -1.30833 & 1.24510 & 0.89&53913.0320 \\ 
N06 & -0.93603 & 1.24717 & 0.87&53913.0601 \\ 
N07 & -0.56239 & 1.24849 & 0.77&53913.0873 \\ 
N08 & -0.18812 & 1.24983 & 0.79&53913.1142 \\ 
N09 & 0.18716 & 1.24878 & 0.83&53913.1628 \\ 
N10 & 0.56203 & 1.24997 & 0.88&53913.1893 \\ 
N11 & 0.93792 & 1.24950 & 0.79&53913.2163 \\ 
N12 & 1.31143 & 1.24874 & 0.79&53913.2428 \\ 
N13 & 1.68723 & 1.24975 & 0.74&53913.2836 \\ 
N14 & 2.06240 & 1.24960 & 0.65&53913.3101 \\ 
N15 & 2.43782 & 1.25014 & 0.98&53913.3366 \\ 
N16 & 2.81285 & 1.25019 & 0.95&53915.1082 \\ 
N17 & -2.81131 & 1.75030 & 1.29&53913.9998 \\ 
N18 & -2.43616 & 1.74961 & 1.27&53914.0263 \\ 
N19 & -2.06159 & 1.74952 & 1.17&53914.0529 \\ 
N20 & -1.68770 & 1.74842 & 1.25&53914.0794 \\ 
N21 & -1.31190 & 1.74986 & 1.21&53914.1062 \\ 
N22 & -0.93705 & 1.75058 & 1.20&53914.1328 \\ 
N23 & -0.56215 & 1.75068 & 0.93&53914.1846 \\ 
N24 & -0.18647 & 1.74973 & 0.88&53914.2115 \\ 
N25 & 0.18823 & 1.74932 & 1.05&53914.2381 \\ 
N26 & 0.55797 & 1.74666 & 1.27&53914.2653 \\ 
N27 & 0.93774 & 1.75029 & 1.18&53915.1678 \\ 
N28 & 1.31297 & 1.74942 & 1.15&53915.3362 \\ 
N29 & 1.68820 & 1.75016 & 0.97&53915.0015 \\ 
N30 & 2.06322 & 1.75033 & 1.18&53915.0280 \\ 
N31 & 2.43796 & 1.75000 & 0.99&53915.0544 \\ 
N32 & 2.81276 & 1.74979 & 0.91&53915.0813 \\ 
  \hline
  \end{tabular}
  \label{tbl:pointingcentres}
\end{table}

\subsection{Data reduction}
The data reduction is carried out using a pipeline developed by the Cambridge Astronomical Survey Unit (CASU), which is specifically designed for processing wide-field mosaic images and is described in detail in \citet{Eduardo2008}. This pipeline was used to process for example the Isaac Newton Telescope Photometric H$\alpha$ Survey (IPHAS, \citeauthor{Drew2005} \citeyear{Drew2005}; \citeauthor{Barentsen2014} \citeyear{Barentsen2014}). In what follows we describe the most relevant steps in the reduction process. For clarity we will refer to fields as the whole field of view of the instrument, and to frames as single CCD images.

First, the standard reduction steps such as bias-subtraction, flat-field correction and sky-background subtraction are performed. Cosmic rays, bad pixels and/or columns are flagged in confidence maps. Source detection and extraction are performed using standard IRAF routines.

\subsubsection{Photometry}
In the next step optimal aperture photometry is performed, using a series of aperture radii to determine the median seeing for each frame. The magnitudes are determined using an aperture radius that corresponds to the seeing, where a photometric correction taking into account the aperture shape and size is included. A radial distortion correction is applied to prevent systematic errors in the photometry due to distortion of the detector response towards the edges of the field. The source extraction is performed taking into account the high source densities by including a crowded field analysis \citep{Irwin1985}.
Magnitudes are calculated in the Vega system using the following formula:\\
\large
\begin{equation}
m = \text{ZP} - 2.5\ \text{log}\left(\frac{\text{flux}}{\text{exptime}}\right) - \text{apcor}
\end{equation}
\\\newline
\normalsize
where flux is the number of counts within the aperture, exptime is the exposure time in seconds, and ZP is the CCD zeropoint per night (corrected for airmass and atmospheric extinction variations), determined using a series of Landolt stars \citep{Landolt1992}. Apcor is a photometric correction accounting for the aperture shape and size. The values of the zeropoints for each of the filters are given in Table \ref{tab:zeropoints}. There are two nights on which no GBS data was taken due to bad weather conditions. If we did not observe standard star fields on a given night, we use an uncertainty of 0.05 mag for the zeropoint of that night. 

Given that there are no standard calibration sources for the H$\alpha$ filter, the H${\alpha}$ magnitudes quoted in the catalogues are calculated using a constant zeropoint ZP$_{H{\alpha}}=30$. 
We will tie these magnitudes to the r$^{\prime}$-band to make them conform to the Vega magnitude scale.
To achieve this, we need take into account the strong absorption feature in the Vega spectrum that lowers the flux below the continuum level \citep{Drew2005}. Assuming that the broadband $r^{\prime} - i^{\prime}$ colour is well-defined, we only apply a correction to the r$^{\prime} - H{\alpha}$ colour index. We correct in such a way that Vega has r$^{\prime} - H{\alpha}$ = 0. To achieve this we tie the zeropoint of the $H{\alpha}$ filter to the r$^{\prime}$-band, and furthermore we correct for the Vega $H{\alpha}$ excess: 
\begin{center}
\large
\begin{equation}
\text{ZP}_{H\alpha} = \text{ZP}_{r^{\prime}} - 3.58
\end{equation}
\end{center}\normalsize
We determined the $H{\alpha}$ excess of Vega by calculating the synthetic magnitudes of an HST spectrum (see Section \ref{sec:synphot}).

\begin{table}
 \begin{minipage}{\columnwidth}
 \begin{center}
  \caption{Photometric zeropoints (in mag) for the nights we observed GBS fields, determined using a series of Landolt stars. In case we did not observe standard fields, we use an uncertainty of 0.05 mag. The MJD is given for the middle of the night, in days. On MJD 53909 and 53910 we did not observe GBS fields due to bad observing conditions.}
  \begin{tabular}{ccccc}
\hline
  MJD & ZP$_{r^{\prime}}$ & $\sigma_{r^{\prime}} $ & ZP$_{i^{\prime}}$ &  $\sigma_{i^{\prime}}$ \\
  \hline\hline
53907 & 25.48 & 0.02 & 24.81 & 0.01 \\
53908 & 25.55 & 0.05 & 24.81 & 0.05 \\
53911 & 25.55 & 0.05 & 24.81 & 0.05 \\
53912 & 25.54 & 0.01 & 24.80 & 0.01 \\
53913 & 25.53 & 0.02 & 24.82 & 0.01 \\
53914 & 25.55 & 0.05 & 24.81 & 0.05 \\
53915 & 25.56 & 0.01 & 24.81 & 0.01 \\
53916 & 25.55 & 0.05 & 24.81 & 0.05 \\
  \hline
  \end{tabular}
  \label{tab:zeropoints}
  \end{center}
 \end{minipage}
\end{table}

For the calculation of the photometric errors we take into account the uncertainty of the zeropoint calculations (Table \ref{tab:zeropoints}), and in addition we add Poissonian errors in quadrature to incorporate photon counting statistics. 

A morphological classification flag is provided based on a comparison of the curve-of-growth of the flux versus aperture radius for each detected object with the curve-of-growth of the stellar locus, which is well-defined and can be used to first approximation to classify objects. Sources that are within 2 - 3 sigma of the stellar locus are generally flagged as stellar, while objects 3 - 5 sigma below (which signifies a sharper point-spread function [PSF]) as noise-like, and those 2 - 3 sigma above (more diffuse PSF) as extended. Based on the ellipticity of the PSF, ambiguous cases are flagged as borderline stellar and borderline extended. Sources that appear saturated in the observations are flagged separately. We also include a flag to indicate if there are bad pixels (e.g. hot or dead pixels) within the aperture of a source entry. This is done using the confidence maps mentioned earlier. This analysis is performed for each frame independently, and later the information is merged together for each field. Table \ref{tbl:flags} shows the possible classification flags associated with the observations. 

\begin{table}
 \begin{minipage}{\columnwidth}
 \begin{center}
  \caption{Morphological classification flags included in the catalogues by analysing the curve-of-growth of the flux versus aperture radius for each object.}
  \begin{tabular}{cc}
\hline
  Flag & Morphology  \\
  \hline
  0     & Noise \\ 
  1     & Extended \\
  -1   & Stellar \\
  -2   & Borderline stellar \\
   -3  & Borderline extended \\
   -7  & Bad pixel(s) in aperture \\
  -9   & Saturated \\
  \hline
  \end{tabular}
  \label{tbl:flags}
  \end{center}
 \end{minipage}
\end{table}
\subsubsection{Astrometry}

The astrometric calibration of the Mosaic-II images is complicated by the significant distortion in the instrument, where the pixel scale decreases by 4 per cent from the center to the edge of the instrument field-of-view. Because some fields contain a large number (more than 1500) of astrometric standards, we can use these to calculate the geometric distortion correction for the Mosaic-II instrument. We compare the absolute positions of the astrometric standards to their pixel positions on the detector, and fit a 4th-order polynomial as a function of pixel position. We use this distortion correction to convert between pixel positions and positions on an undistorted meta frame. 

We use astrometric standards from the second version of the USNO CCD Astrograph Catalog (UCAC2; \citeauthor{Zacharias2004} \citeyear{Zacharias2004}) to match against stars on each of the 10s r$^{\prime}$-band images. Stars that were saturated, blended or did not appear stellar were removed, and the centroids of the remaining stars were measured and corrected for distortion. An astrometric solution for each image was determined by fitting a position offset and a four parameter transformation matrix between the observed distortion-corrected positions and the catalogued position of the astrometric standards. Outliers were iteratively removed until the solution converged.

Due to the large changes in stellar density over the GBS fields, the number of UCAC2 standards coinciding with stars on each image varied from 50 in low density regions up to over 1500 in high density fields. On average some 390 UCAC2 standards were used to determine the astrometric solution of each image. The mean root-mean-square (rms) residual of the solution is 0.076 arcsec in right ascension and 0.065 arcsec in declination, with standard deviations of 0.011 arcsec in both coordinates. 

\begin{center}
\begin{figure}
  \includegraphics[width=8.5cm, keepaspectratio]{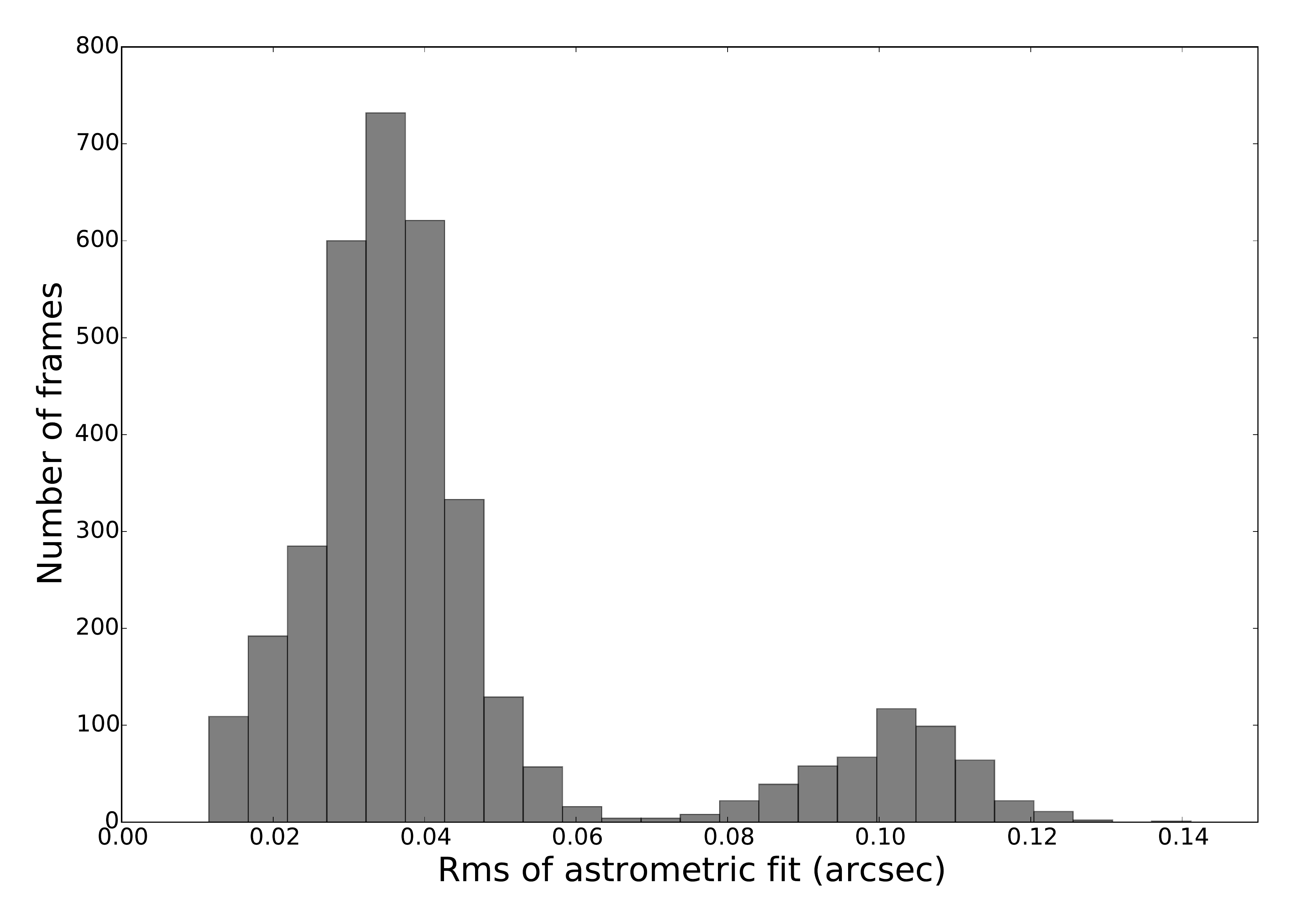}
  \caption{Distribution of the rms residuals (1$\sigma$) of the astrometric solutions for each of the frames. The average value is 0.044 arcsec. For the frames with an rms higher than $\sim$ 0.07 arcsec there were only a limited amount of astrometric standard stars available.}
  \label{fig:rmsdistribution}
\end{figure}
\end{center}

Next, stars on the 10s images were used to create a secondary astrometric catalogue to transfer the astrometric solutions to the deep r$^{\prime}$, i$^{\prime}$ and H$\alpha$ images. The same iterative procedure used for the 10s images was applied to determine the astrometric solution of the deep images. Typically, a large number of secondary astrometric standards (100 to over 3500) were available. The solutions have average rms residuals of 0.032 arcsec in right ascension and 0.030 arcsec in declination.
The distribution of the 1$\sigma$ rms values for each frame is shown in Figure \ref{fig:rmsdistribution} and has a mean value of 0.044 arcsec. 
The bimodal distribution of the rms values is due to a bimodal distribution of the number of standards available for calculating the astrometric solution. 
Inspection of the astrometric positions shows that there is a population of sources at the bright end of the magnitude distribution (i$^{\prime}\leq$16 and flagged as saturated) for which the position changes by more than the optical astrometric uncertainty between the three filters. This is caused by the PSF centering algorithm having trouble finding the position of the peak of the PSF. We expect the astrometry of saturated sources to be less accurate, so the position of the peak may change by up to a few arcseconds (depending on the magnitude). However, these sources also subtend several arcseconds on the CCD, so there is never the danger of source confusion or mismatching with other objects.

\subsubsection{Band-merging}
In a last step the source catalogues in different filters are merged to generate one catalogue per field which contains for each source the position, r$^{\prime}$, i$^{\prime}$ and H$\alpha$ magnitudes and their errors, and a morphological classification flag for each band (together with extensive auxiliary information that can be found in the catalogues). The band merging is performed using positional information only. It is driven from the r$^{\prime}$-band photometry, and offsets to the closest sources in the i$^{\prime}$ and H$\alpha$ observations are computed with respect to a common field centre and corrected for. The matching is performed between these corrected coordinates with a search radius of up to 2.5 arcseconds, and the conversion to sky coordinates is performed using the unique r$^{\prime}$-band reference frame. The search radius of 2.5 arcsec (much larger than the astrometric uncertainties) is motivated by the astrometric uncertainties for bright (saturated) stars. If a source is present in the r$^{\prime}$-band, these coordinates are quoted in the merged catalogue. If there is no source in r$^{\prime}$, we use the corrected coordinates of the i$^{\prime}$-band position to convert to sky coordinates (again using the r$^{\prime}$-band reference frame). The result is one merged catalogue for the original field as well as for the offset observations of the same field.

\subsection{The optical catalogue in numbers}
The data products generated using this pipeline consist of the reduced mosaic images and the derived object catalogues. The catalogues are stored in multi-extension FITS files as binary FITS tables consisting of a set of descriptors for each detected object. Each catalogue header contains a copy of the relevant telescope FITS header content in addition to detector-specific information.
The resulting optical catalogue of the GBS area contains positions and magnitudes of about 22.5 million objects detected in one or more bands at the 5$\sigma$ level. An example of the most relevant catalogue entries is shown in Table \ref{tab:optcat}. In addition to these tables, the full single-band catalogues as well as the merged catalogues are available in electronic form at Vizier (http://vizier.u-strasbg.fr).
Table \ref{tab:catalognumbers} lists some numbers that characterise the catalogue. 
\begin{table*}
 \centering
 
  \caption{Five entries of the most relevant information in the optical catalogue, including the position, magnitudes and morphological classification of objects. The positional uncertainties are the rms residuals from the astrometric fit to UCAC2 sources in the same field. The photometric measurements are quoted in magnitudes. The photometric uncertainties include the zeropoint uncertainty and uncertainties due to photon counting statistics. }
  \begin{tabular}{cccccccccccc}
  \hline
 RA ($^{\circ})$ & Dec $(^{\circ})$  &  rms ($^{\prime\prime})$ & $r^{\prime}$ & $\sigma_{r^{\prime}}$ &  $i^{\prime} $ & $\sigma_{i^{\prime}}$ & $H{\alpha}$ & $\sigma_{H{\alpha}}$ &$r^{\prime}$ flag & $i^{\prime}$ flag &$H{\alpha}$ flag \\
  \hline\hline
265.392822&	-26.388750&	0.03&	18.10	&0.01	&16.65	&0.01&	17.51	&0.14&-1&	-1	&-1\\
265.624083&	-26.288745&	0.03&	18.09	&0.01	&15.41	&0.01&17.19	&0.14&-1&-9&-1\\
265.591368&	-26.353128&	0.03&	18.10	&0.01	&16.91	&0.01&17.46	&0.14&-1&	-1&	-1\\
265.682983&	-26.335877&	0.03&	18.10	&0.01	&17.04	&0.01&	17.63	&0.14&-1&	-1	&-1\\
265.495081&	-26.349439&	0.03&	18.10	&0.01	&16.09	&0.01&	17.41	&0.14&-1&	-1	&-1\\
  \hline
  \end{tabular}
  \label{tab:optcat}
\end{table*}

\begin{table}
 \centering
  \caption{Statistics of the optical catalogue presented in this work, quoted in units of millions. The numbers given here refer to objects detected only in a certain filter with a certain flag (they do not take into account flags in other filters).}
  \begin{tabular}{c|ccc}
Filter & N$_{r^{\prime}}$ ($\times 10^6$)& N$_{i^{\prime}}$ ($\times 10^6$)& N$_{H\alpha}$ ($\times 10^6$) \\\hline
 Stellar & 11.82 & 15.37 & 11.41 \\
 Probable stellar & 0.74 & 0.62 & 1.02 \\
 Extended & 3.66 & 3.47 & 3.44 \\
 Probable extended & 0.57 & 0.32 & 0.67 \\
 Saturated & 0.32 & 0.97 & 0.08 \\
 \hline
 Total & 17.11 & 20.75 & 16.62 \\
  \hline
  \end{tabular}
  \label{tab:catalognumbers}
\end{table}

\begin{center}
\begin{figure}
  \includegraphics[width=9cm, keepaspectratio]{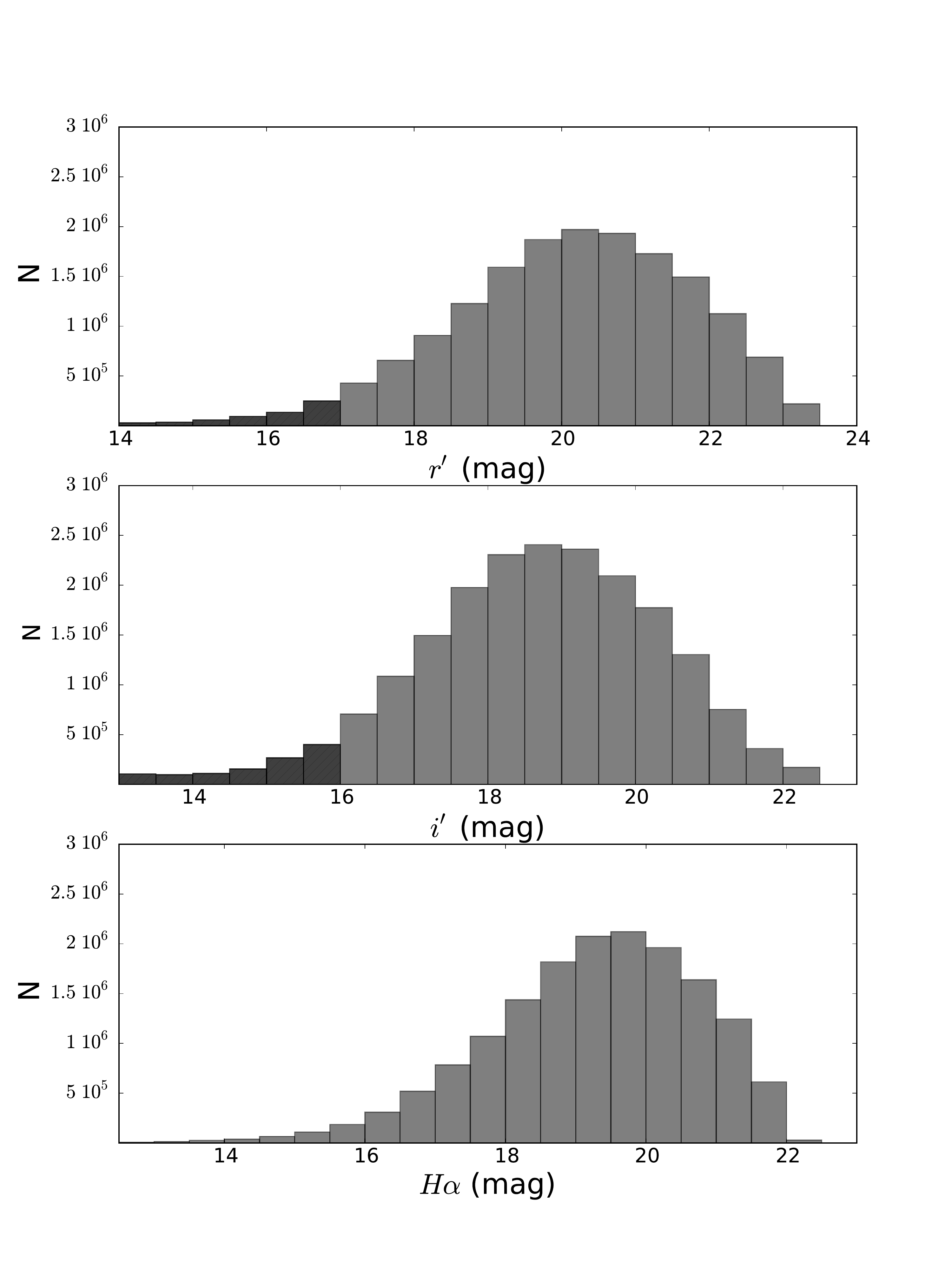}
  \caption{Distribution of the magnitudes of all objects that are detected as stellar in all three bands within the sensitivity limits of the data. The dark grey areas indicate the magnitude range where sources are saturated. The photometry for these objects is therefore uncertain, and these magnitudes should be interpreted with care. The average 5$\sigma$ detection limit of the observations is r$^{\prime}$ = 22.5, i$^{\prime}$ = 21.1.}
  \label{fig:rihadist}
\end{figure}
\end{center}

\begin{figure}
  \includegraphics[width=9cm, keepaspectratio]{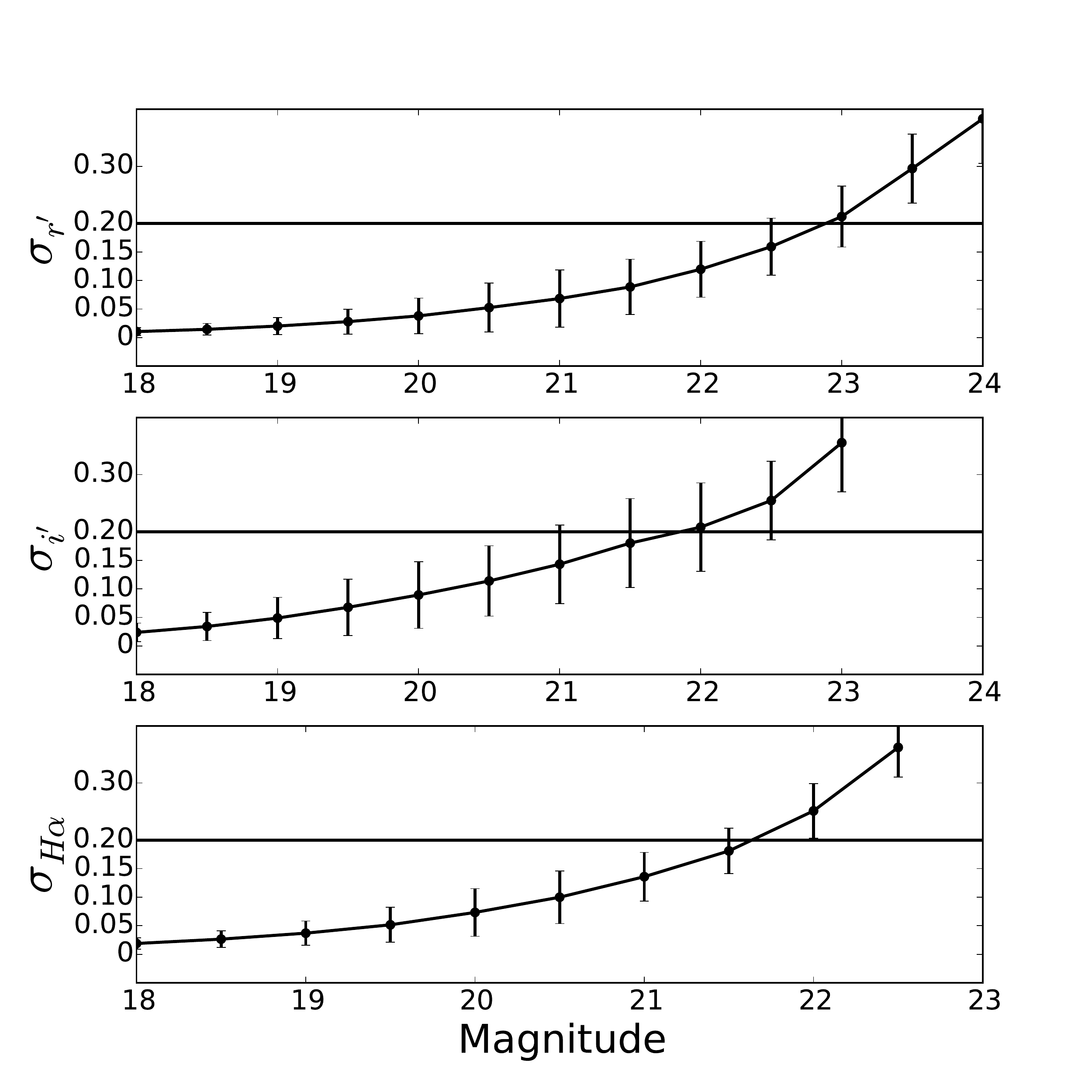}
  \caption{Photometric (Poissonian) uncertainties as a function of magnitude. Systematic uncertainties are not included. The errorbars represent the scatter on the mean magnitude in each bin. The horizontal solid line represents the 5$\sigma$ detection limit. The curved solid line only connects the datapoints.}
  \label{fig:poissonphot}
\end{figure}

The distribution of the r$^{\prime}$, i$^{\prime}$ and H${\alpha}$ magnitudes are shown in Figure \ref{fig:rihadist}. In Figure \ref{fig:poissonphot} we show the mean Poissonian photometric uncertainties of the catalogue as a function of magnitude. The errorbars represent the scatter on the mean magnitude. The horizontal line represents the 5$\sigma$ detection limit. The mean 5$\sigma$ depth of the observations is r$^{\prime}$ = 22.5 and i$^{\prime}$ = 21.1 but depends on the seeing. Our catalogue is complete  down to r$^{\prime}$ = 20.2 and i$^{\prime}$ = 19.2 mag at the 5$\sigma$ level, although we make no attempt to quantify the detection probability near bright objects. These completeness limits are the brightest 5$\sigma$ detection limits in the optical catalogue.

\subsection{Detection bias around bright stars}
\label{subsec:brightsources}
It is plausible that the presence of a large number of saturated sources affects the detection rate of faint stars in their vicinity. In this paragraph we look for evidence of such a detection bias in our data.

As our optical observations cover an area on the sky that is largely devoid of X-ray sources, we can compare the population of optical sources in the vicinity of X-ray sources with a population of sources in randomly selected positions in the GBS survey area.
For each X-ray source we generate a set of 10 randomly selected offset
positions (excluding the X-ray error circle). Around each offset
position, we determine the distribution of the magnitudes and the
distances between the X-ray position and the location of the stars. We
consider stars up to a distance equal to the radius of 4R$_{\sigma}$(where R$_{\sigma}$
is the 1$\sigma$ astrometric error circle of the X-ray observations). So, we
compare similarly sized areas on the sky. We make sure that there is no
overlap with the 4R$_{\sigma}$ area around the X-ray position. We furthermore
restrict our offset positions to be within 200 arcsec of the X-ray
position to minimise the effect of extinction variations across the
GBS sky area.
We create histograms for the 10 sets of 1640 offset positions and average them, and denote this the mean histogram H$_{mag, mean}$. We create the same histogram of magnitudes for the optical sources in the X-ray error circles, denoted H$_{mag, X}$, and for an additional independent offset position H$_{mag, random}$.

\begin{center}
\begin{figure}
  \includegraphics[width=7.5cm, keepaspectratio]{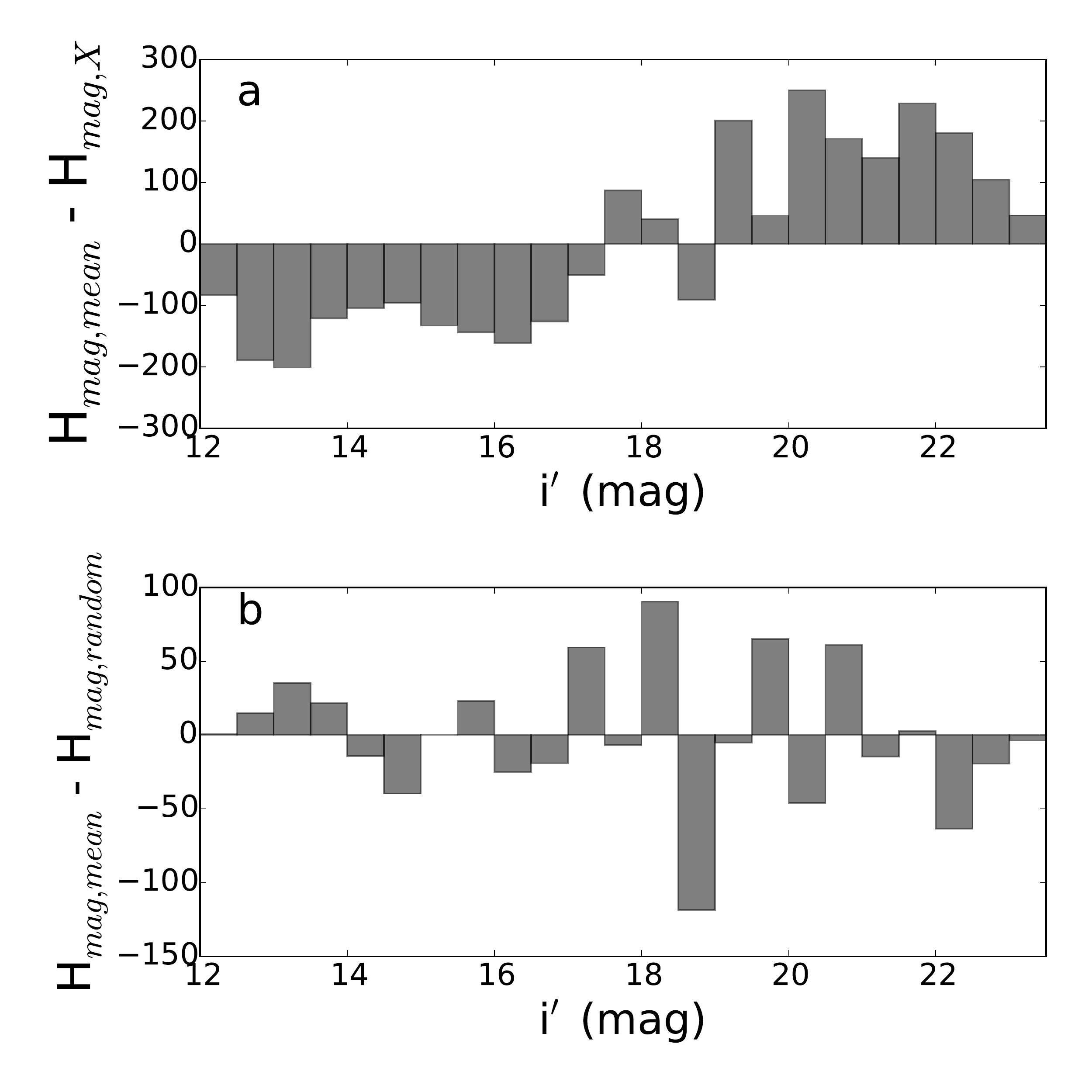}
  \caption{Top: histogram of the average number of optical sources in offset positions (H$_{mag, mean}$) minus the number of optical sources present in the X-ray error circle (H$_{mag, X}$) as a function of magnitude. There is an excess of bright sources in the X-ray error circles with respect to the average number of bright optical sources in regions which do not include the X-ray positions. Bottom: same as the top, but now we have replaced the number of sources found in the X-ray error circles with the number of sources found in independent offset fields (H$_{mag, random}$). This difference varies around zero, which is expected for random (independent) positions on the sky.}
  \label{fig:diffhistmag}
\end{figure}
\end{center}

Panel a of Figure \ref{fig:diffhistmag} shows the difference between H$_{mag, mean}$ and H$_{mag, X}$, while panel b of the same figure shows the difference between H$_{mag, mean}$ and H$_{mag, random}$.
The two main features of Figure \ref{fig:diffhistmag}a (H$_{mag, mean}$ - H$_{mag, X}$) are the negative and positive values. Negative values indicate an excess of bright sources in the X-ray error circles with respect to the average. Clearly bright optical sources have a preference for residing in the neighbourhood of an X-ray source, while for fainter magnitudes there is no such apparent preference. We see that the presence of bright sources affects the detection of fainter sources in error circles of X-ray positions, indicated by the excess of faint sources when compared to the average (the positive values in panel a of Fig. \ref{fig:diffhistmag}). This effect can also be observed on the processed images, specifically the r$^{\prime}$-band images for which we have one short 10s exposure and a long exposure. Multiple sources can be resolved around bright objects in the short exposures because there is no CCD blooming around them. In the long exposures the charge leaks into adjacent pixels and faint objects can no longer be detected.
To illustrate that the effect visible in the difference histogram is real, we also show H$_{mag, mean}$ - H$_{mag, random}$ (Fig. \ref{fig:diffhistmag}b). We see that in this case the difference fluctuates around zero, as is expected for random locations on the sky. 

Repeating the same analysis described above, but excluding error circles that contain bright sources, we find that the apparent lack of faint sources within the X-ray error circles disappears, indicating that the observed effect is linked to the presence of bright stars. We conclude that there is a detection bias: the detection efficiency for faint sources is lower in the vicinity of bright objects.

\begin{center}
\begin{figure}
  \includegraphics[width=7.5cm, keepaspectratio]{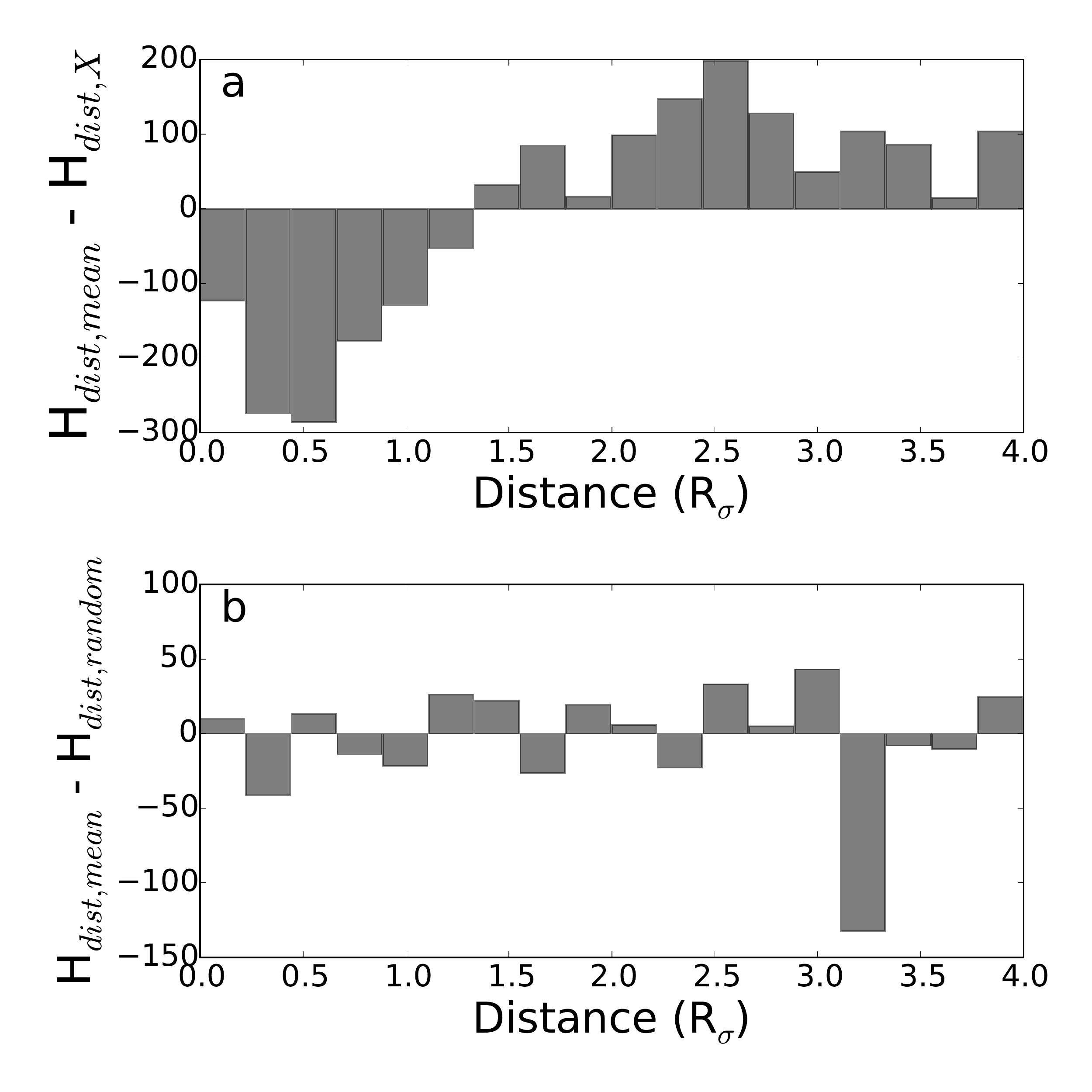}
  \caption{Same as Fig. \ref{fig:diffhistmag}, instead we show the distribution of the number of sources as a function of distance from the best-fit X-ray position. The top panel shows H$_{dist, mean}$ - H$_{dist, X}$. We see an excess of sources towards the centre of the X-ray error circles with respect to the average of offset positions in the sky. The bottom panel shows H$_{dist, mean}$ - H$_{dist, random}$, and we see variations around zero as expected for random positions on the sky.}
  \label{fig:diffhistdist}
\end{figure}
\end{center}

Panels a and b of Figure \ref{fig:diffhistdist} show the same as in Fig. \ref{fig:diffhistmag}, but instead of the stellar magnitudes, we use the distance from the centre of the error circle (normalised to R${_\sigma}$) and we sum over all source magnitudes. In this case we see negative values for offsets smaller than $\sim$ 1.5$\sigma$, indicating that the optical sources have a preference for residing near the centre of the error circle. This result indicates that we are finding real counterparts to the X-ray sources. This does not mean, however, that all matches we find at distances larger than 1.5$\sigma$ are automatically interlopers. In fact, from our assumption (that the distribution of true X-ray positions within the error circles can be described by a Rayleigh distribution, see Section \ref{sec:likelihood}), we expect that $\sim$ 30 per cent of the candidate counterparts will be located outside this region. 

\subsection{Synthetic photometry}
\label{sec:synphot}
We compare our observations to synthetic photometry using all the available optical data. 
We consider synthetic photometry of solar-metallicity main-sequence (MS) and giant stars using stellar SEDs from the Pickles library \citep{Pickles1998}. The binning of these spectra is sufficiently small (5$\AA$) that we can use them to compute synthetic photometry for our r$^{\prime}$ and i$^{\prime}$ filters as well as for the narrow-band H$\alpha$ filter. We recompute the grids of these filter profiles to match the binning of the spectra, meaning that for each spectral bin we compute the filter transmission value at the midpoint of the bin.
We define the synthetic colours in the Vega system as
\begin{center}
\large
\begin{equation}
m_1 - m_2 = -2.5 \text{log } \frac{\int T_{1,\lambda} F_{\lambda} d\lambda}{\int T_{1,\lambda} F_{\lambda, \text{V}} d\lambda} + 2.5 \text{log} \frac{\int T_{2,\lambda} F_{\lambda} d\lambda}{\int T_{2,\lambda} F_{\lambda, \text{V}} d\lambda}
\end{equation}
\normalsize
\end{center}
where the filter transmission profiles are labeled $T_x$ (see Fig. \ref{fig:transmissionprofiles}), $F_{\lambda}$ is the synthetic spectrum per spectral type and $F_{\lambda, \text{V}}$ is the spectrum of Vega \citep{Bohlin2007}.
We calculate the positions of MS stars in a synthetic colour-colour diagram (CCD) for spectral types ranging from O5V to M5V. For the giants we use spectral types from O8III to M5III. To comply with a Vega-based magnitude scale, we normalise the synthetic colours with respect to an HST spectrum of Vega.

To compare the simulated r$^{\prime} - H{\alpha}$ colours with observations, we calculate these synthetic colours for a range of reddening values using a standard mean Galactic extinction law R = 3.1 \citep{Cardelli1989}.
\begin{figure*} 
  \includegraphics[height=9cm, keepaspectratio]{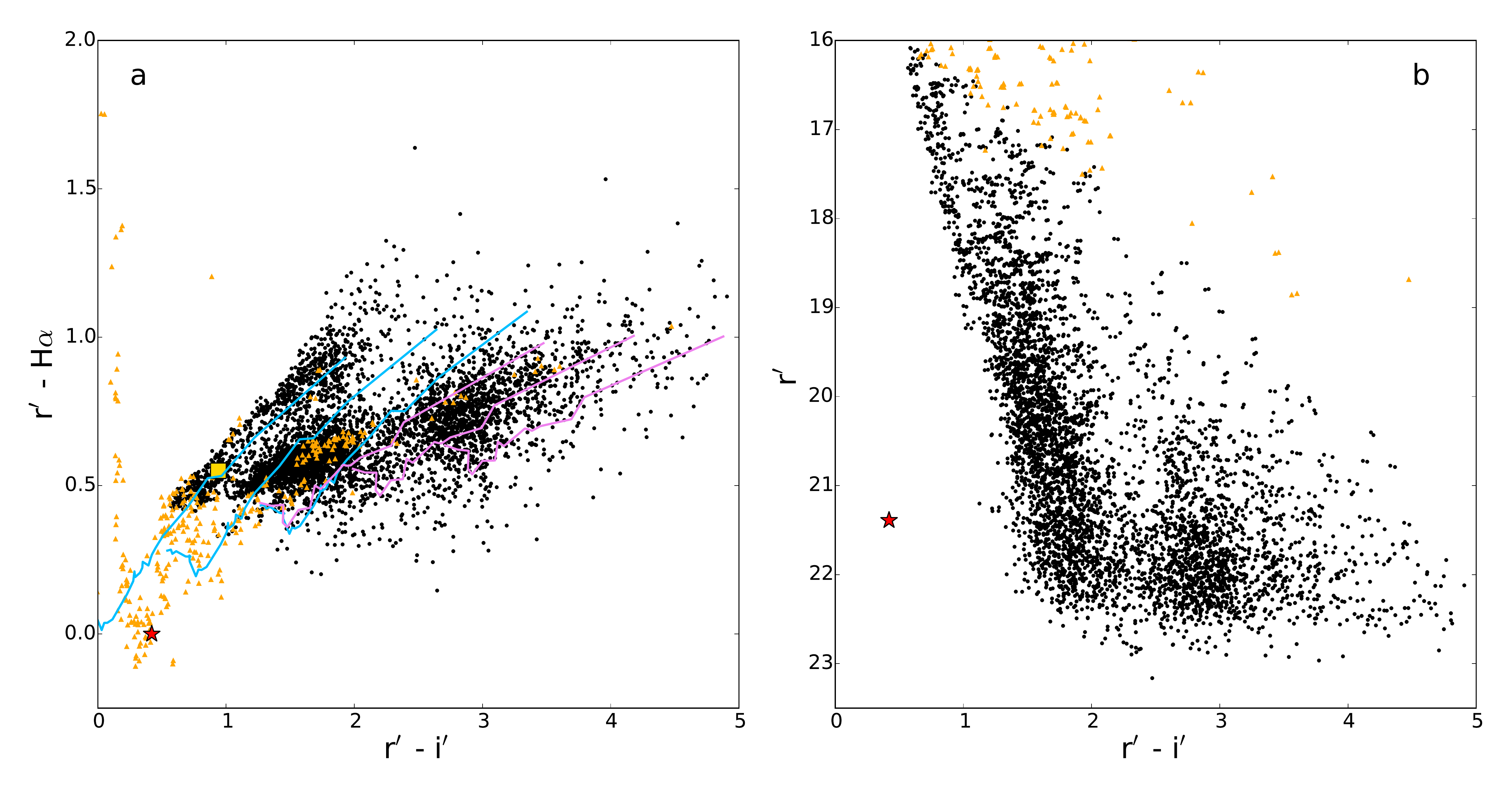}
  \caption{Panel a: comparison between synthetic reddening tracks, with E(B-V) ranging from 0 to 2 for MS (blue) and 2 to 4 for giants (purple), and our observations of field S02. Saturated sources are plotted as orange triangles. Note that we did not shift the synthetic tracks to fit the unreddened MS or the most reddened giant branch. The red star marks an optical counterpart candidate to an X-ray source. The yellow square indicates that an optical counterpart candidate was found, but it is saturated in our observations. The candidate at r$^{\prime} - H{\alpha}$ $\sim 0$ is likely a white dwarf because of the blue colours and indications for extreme H$\alpha$ absorption features. Panel b: CMD of the same field. The MS and giant branches can be clearly identified in this diagram. The candidate counterpart has a very blue colour when compared to other stars in the field.}
  \label{fig:syntrackS02}
\end{figure*}

\subsection{Comparison of simulated and observed data}
\label{sec:comparisonsimobs}
Figure \ref{fig:syntrackS02} shows the colour-colour and colour-magnitude diagrams (CMD) of field S02, with the synthetic MS and giant tracks overplotted. 

The observed unreddened MS track falls nicely along the theoretical track. We have plotted the MS tracks for values of E(B-V) = [0,1,2], while we show giant tracks with E(B-V) = [2,3,4]. By calculating these reddening tracks across spectral types for a range of E(B-V) the locus of the theoretical MS shifts towards slightly higher ($ r^{\prime} - H{\alpha}$) and redder ($r^{\prime} - i^{\prime}$) colours. Panel a of Fig. \ref{fig:syntrackS02} shows the colour-colour diagram of this field. We see the locus of unreddened MS stars (coinciding with the E(B-V) = 0 synthetic track) and a locus of reddened stars which are likely giants at higher $r^{\prime} - i^{\prime}$. These giants are intrinsically more luminous, so we can detect them at larger distances. The redder colours (with respect to the unreddened MS) are a combination of intrinsic colour and interstellar reddening. In the CMD (Fig. \ref{fig:syntrackS02} panel b) we can recognise the same two branches of MS stars and giants. The red star marks the candidate counterpart to an X-ray source (see Section \ref{sec:xmatch} for more details). The yellow square indicates a saturated optical counterpart candidate. From the CMD we see that one of them is a very blue object compared to the MS stars in the field, and the CCD shows that it is an outlier showing signs of an extreme H$\alpha$ absorption line compared to the rest of the field stars (inferred from the comparatively low r$^{\prime} - H{\alpha}$ colour index). The combination of a blue system with potential H$\alpha$ absorption suggests that this object could be a white dwarf counterpart to an X-ray source.
It should also be noted that above $r^{\prime} \sim 17$, virtually all sources are flagged as saturated (regardless of their colour). These sources are plotted as orange triangles and the photometric information of these optical counterpart candidates is unreliable.

\subsection{Effect of reddening on the observed colour-colour diagrams}
We can construct ($ r^{\prime} - H{\alpha}, r^{\prime} - i^{\prime}$) CCDs and ($r^{\prime}, r^{\prime} - i^{\prime}$) CMDs along different lines of sight to gauge the effect of different amounts of reddening on the observables and different stellar populations. We publish the scriptsfor making these plots together with the catalogues.

\begin{figure} 
  \includegraphics[height=6cm, keepaspectratio]{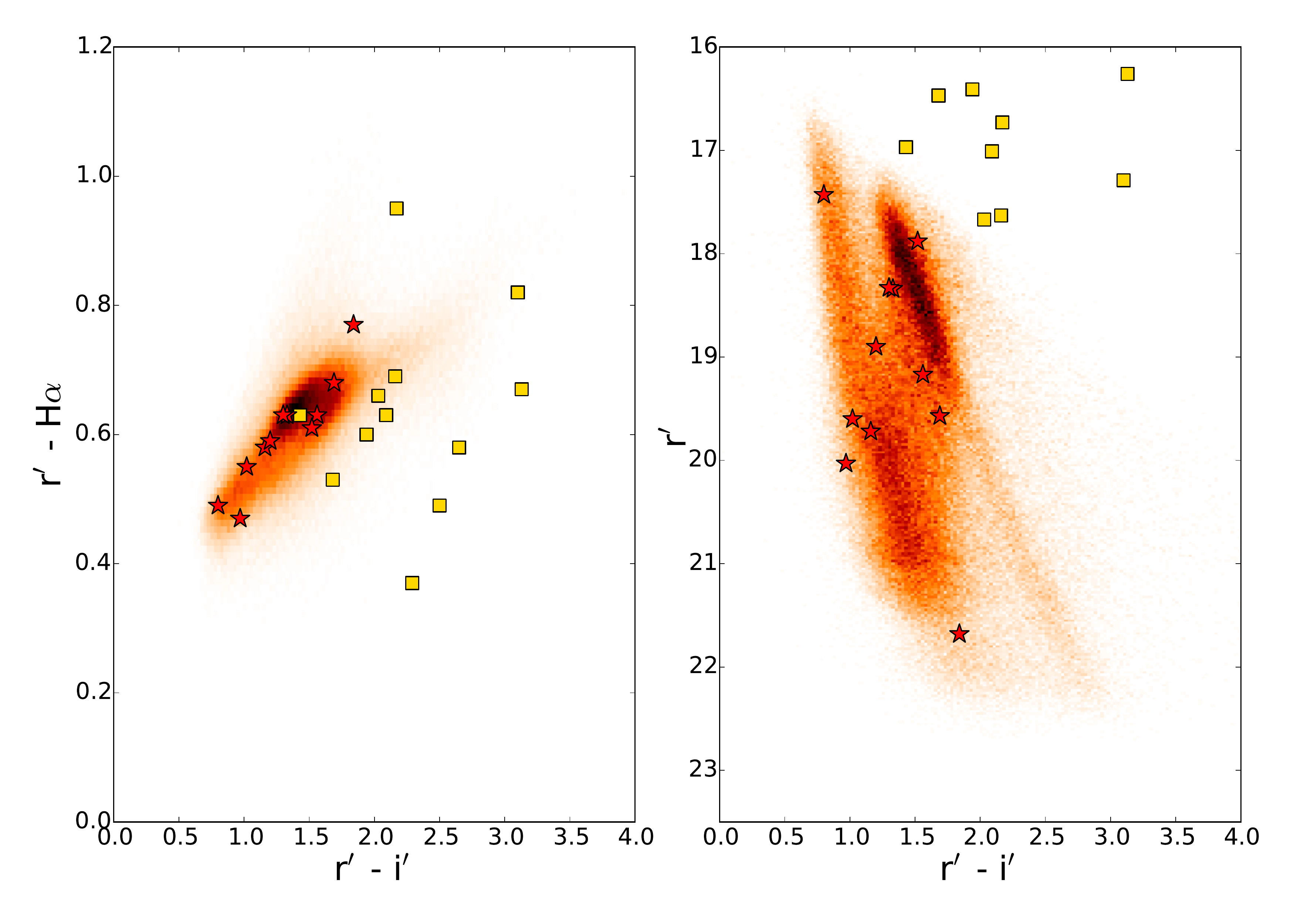}
  \caption{2D histograms of the CCD and CMD of field S15, located on the outskirts of the Bulge in an area of low dust extinction. We used a binsize of 0.025 by 0.025 mag. The low reddening (A$_{r^{\prime}} \sim$ 3.3, see text) towards this field results in very well populated MS and giant branches. Candidate optical counterparts to X-ray sources are marked with red stars and yellow squares. Yellow squares are those candidates that are flagged as saturated. The total number of sources towards this field is about 320000.}
  \label{fig:fieldS15}
\end{figure}

\begin{figure} 
  \includegraphics[height=6cm, keepaspectratio]{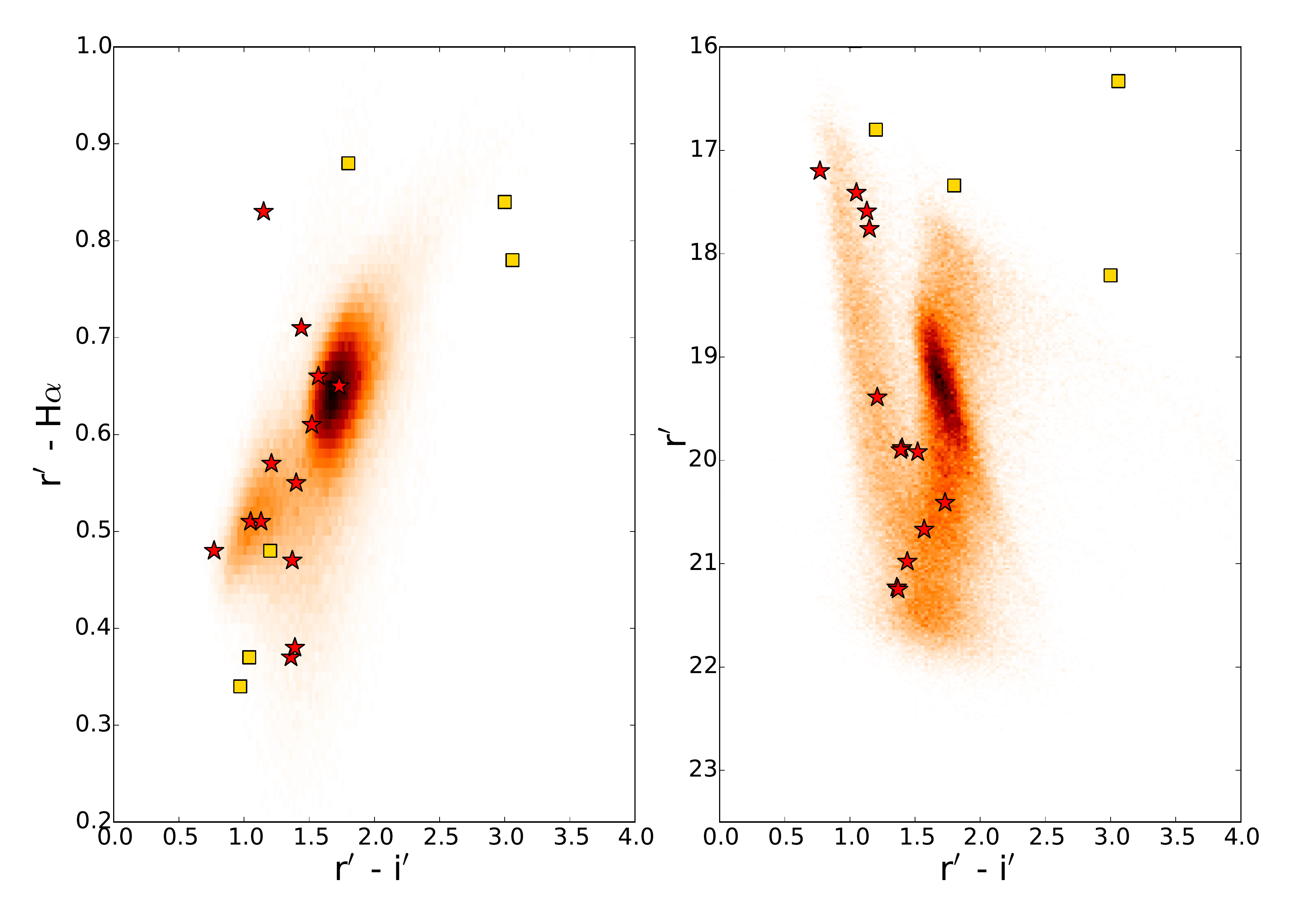}
  \caption{Same as Fig. \ref{fig:fieldS15} for field N30 suffering from higher reddening (A$_{r^{\prime}} \sim $ 4.2, see text). The increased extinction results in the MS track having less stars relative to the giant branch. Moreover the populations move to fainter magnitudes and redder $r^{\prime} - i^{\prime}$ colours.}
  \label{fig:fieldN30}
\end{figure}

In Figure \ref{fig:fieldS15} we show the CCD and CMD of field S15. 
This field is located on the outskirts of the Bulge, and has a lower reddening along the line of sight than the median GBS fields. We can distinguish between the MS track (left) and the giant branch (right) in the CMD, though the separation between the two is not strict. Because of the low extinction we see MS and giant stars out to large distances, resulting in two well populated branches. The total number of sources in this field is $\sim$ 320000 objects. The stars in the gap between the branches are likely reddened MS stars that are located in between the foreground MS stars and giant stars further out. We can get a rough estimate of the reddening towards the field using the CCDs. We assume that the two observed loci of stars are two different populations of stars, and that the left-most clump is comprised of foreground MS stars and the right clump consists of typically red clump (RC) stars towards the Bulge. These RC stars are intrinsically bright and approximately standard candles, and are generally assumed to trace the population of Galactic Bulge stars. We assume M$_V$ = 1 and B - V = 1 for late-type RC stars \citep{Bilir2013}, which we combine with the colour transformation from \citet{Jester2005} to obtain an absolute magnitude of M$_{r^{\prime}}$ = 0.7. From the CCD we estimate that these RC stars are reddened by E($r^{\prime} - i^{\prime}$) $\sim$ 1.1, corresponding to A$_{r^{\prime}}$ $\sim$ 3.3 with respect to synthetic tracks of unreddened MS stars (see Fig. \ref{fig:syntrackS02} panel a). 
In the figure we have omitted saturated sources for clarity. Red stars indicate most likely counterparts of the X-ray sources in this field; yellow squares are those most likely counterparts that are flagged as saturated in one or more filters.
If we compare field S15 with an area with higher extinction (Fig. \ref{fig:fieldN30}, field N30), we see that the number of stars in the MS branch decreases, but the giant branch remains similarly well populated, although the giants also move to fainter magnitudes and redder colours due to increased interstellar extinction. This field contains roughly 270000 sources.
Similarly to the previous field, we estimate that the RC stars are reddened by E($r^{\prime} - i^{\prime}$) $\sim$ 1.4, corresponding to A$_{r^{\prime}}$ $\sim$ 4.2. 
The righthand panel of this figure illustrates the typical separation (in $r^{\prime} - i^{\prime}$) between the MS branch and the giant branch. The MS track most likely consists of unreddened foreground stars. Notice that there are more red objects (giants) relative to unreddened objects in this sample. The reddened MS stars in the gap between the 2 branches in Fig. \ref{fig:fieldS15} have largely disappeared. The higher extinction renders them too faint for our survey observations.

\begin{figure} 
  \includegraphics[height=6cm, keepaspectratio]{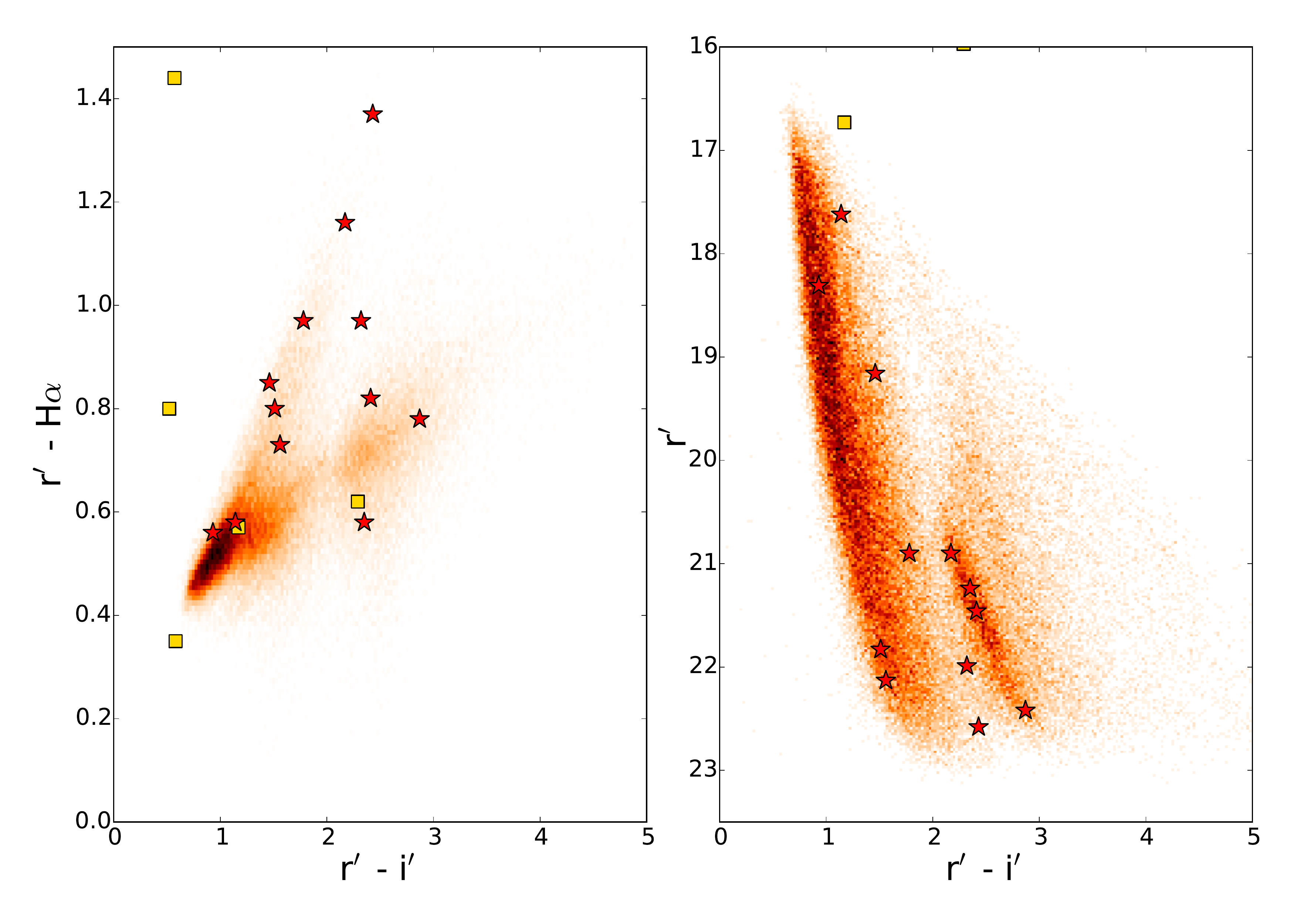}
  \caption{Same as Fig. \ref{fig:fieldS15}, for field S21 close to the Galactic Centre. We estimate the total reddening along the line of sight to be A$_{r^{\prime}}$ $\sim$ 7.8. The total number of stars in this field is about 120000, about a factor of 3 lower than in field S15, due to the high dust extinction.}
  \label{fig:fieldS21}
\end{figure}

Increasing the reddening along the line of sight even more results in yet another structure in both diagrams. In Figure \ref{fig:fieldS21} we show as an example field S21, located close to the Galactic Centre and suffering from high dust reddening. The total number of sources in this diagram has dropped to 120000. 
Following a similar reasoning as given for the other two fields, we estimate a total A$_{r^{\prime}}$ $\sim$ 7.8 for the RC stars observed at $r^{\prime}$ = 22.
We also note that the difference in the number of sources among the 8 frames in this field is about a factor of 2, indicating that there are reddening variations on $\sim$ 10$^{\prime}$ (or smaller) scales along this line of sight.

The CCD and CMD are a powerful tool to search for candidate extreme H$\alpha$ emission and absorption line objects, white dwarfs, carbon stars and various other types of peculiar objects. This is outside the scope of the current paper, but will be performed in the future.


\section{Optical counterparts to GBS X-ray sources}
\label{sec:xmatch}
We now set out to find the optical counterparts to GBS X-ray sources by cross-matching the X-ray positions with three different subsamples of our catalogue. Because of the high source densities, often there is more than 1 possible optical counterpart within the X-ray error circle, and we need to quantify the probability of a source being there by chance. We perform this analysis in a statistical way with two main questions in mind. The first one (Section \ref{sec:fap}) concerns the properties of the X-ray sample as a whole: what is the fraction of optical counterparts that is expected to be present due to chance alignment? 

The other question we want to answer is motivated by the limited spectroscopic resources that are available for source classification (Section \ref{sec:likelihood}). For this we want to address the issue of, given an ensemble of potential optical counterparts to a particular X-ray source, which optical source within the error circle is most likely to be the real one. Addressing these questions requires two different methods, as one question is related to the X-ray sample as a whole while the other question is related to single X-ray sources. 

\subsection{Optical sample selection}
We first apply selection criteria to our optical catalogue in order to eliminate spurious sources, detector artifacts and other sources of possible contamination in our optical sample. 
We apply 3 sets of selection criteria, where the goal is to compare how the cross-matching with the X-ray positions behaves for different optical samples. 

As a baseline we define the most conservative optical selection criteria, where we only select sources that are detected as stellar at the 5$\sigma$ flux level in all three filters (requiring $\sigma_{r^{\prime}}$, $\sigma_{i^{\prime}}$, $\sigma_{H{\alpha}} \leq 0.2$). Sources with non-detections in one or more filters are excluded. We use the morphological classification to discriminate spurious sources (e.g. cosmic rays, detector cross-talk) from extended and point-like objects. Because the goal of this paper is to find the optical counterparts to X-ray sources, we also include sources that were flagged as saturated in our analysis. For these sources the morphological classification has failed, but nevertheless they comprise a large fraction of counterparts and we include them in our analysis. Additionally, we limit the cross-match radius to 0.5 arcseconds between observations in different filters as a criterion to eliminate possibly spurious matches. This selection procedure reduces the total amount of objects for cross-matching to 8.3 million (point) sources:

\begin{enumerate}[i]
  \item $\sigma_{r^{\prime}}$, $\sigma_{i^{\prime}}$, $\sigma_{H{\alpha}} \leq 0.2$
  \item $r^{\prime}$, i$^{\prime}$, H${\alpha}$ != 0
  \item Flag($r^{\prime}$, i$^{\prime}$, H${\alpha}$) = -1 or -9
\end{enumerate}

The second set of selection criteria no longer requires a detection in $H\alpha$. This is motivated by the fact that the images are not as deep in this filter compared to the broadband observations. The exposure time for $H\alpha$ was increased by a factor of 3 with respect to r$^{\prime}$, but Figure \ref{fig:transmissionprofiles} shows that the total area under the $r^{\prime}$ and $i^{\prime}$ filter profiles is more than a factor of 3 larger, resulting in deeper images (as can also be seen from, for example, the number of saturated sources in Table \ref{tab:catalognumbers}). The sample we obtain in this way contains 10.43 million sources detected in $r^{\prime}$ and $i^{\prime}$.\\
In a third sample we also allow optical sources that are detected only in the $i^{\prime}$-band, as it is expected that (due to the high dust extinction towards the Bulge) a significant number of counterparts are heavily reddened, hence these sources will be detected in $i^{\prime}$ but not in $r^{\prime}$. This gives us a sample of 16.34 million optical sources. The large increase of sources in the final sample indicates that the $i^{\prime}$-band images are significantly deeper than the $r^{\prime}$-band observations. We refer to this optical sample as the least conservative one in the rest of the article. 

We note that in the resulting photometry, objects that are brighter than $r^{\prime} \leq 17$ or $i^{\prime} \leq 16$ are generally flagged as saturated. We indicate this in figures showing magnitudes as a hatched area throughout the article.

\subsection{Astrometric uncertainties}
Next, we consider the combined astrometric uncertainties of the X-ray and optical source catalogues. This will determine how large an area around the X-ray source we will consider for identifying sources as optical counterparts to X-ray sources. The choice of error circle is largely motivated by the questions we are trying to answer, and is not the same for our two respective questions.

We start by taking the uncertainties of the Chandra X-ray observations as given in \citet{Jonker2014}. These uncertainties are a function of the number of detected counts and the off-axis detection angle and have been calibrated using the method described in \cite{Evans2010}.
In addition, we take into account the 95 per cent confidence level (CL) for the spacecraft pointing, which amounts to 0.7 arcsec\footnote{http://cxc.harvard.edu/cal/ASPECT/celmon/}. Furthermore, \citet{Primini2011} find residual offsets when comparing the Chandra source catalogue positions to SDSS. They suggest that a component of 0.16 arcsec (1$\sigma$) should be included in the Chandra positional uncertainty. The total 1$\sigma$ uncertainty of the X-ray positions is then:\\\large
\begin{center}
\begin{equation}
R_{\sigma} = \sqrt{(0.4085 \times P)^2 + (0.4085 \times 0.7^{\prime\prime})^2 + 0.16^{\prime\prime 2}}\ \text{arcsec}
\end{equation}
\vspace{1mm}
\end{center}\normalsize
where we have introduced a factor 0.4085 to convert from a 95 per cent CL to a 1$\sigma$ uncertainty. We will use the 95 per cent error circle R$_{95}$, i.e. the radius within which there is a 95 per cent chance to find the X-ray source (which is roughly 2.45$\sigma$ for a Rayleigh distribution).

The uncertainty P is a function of the number of counts, C, and the off-axis angle $\theta$ in arcminutes \citep{Evans2010}:\\
\begin{center}\vspace{-3mm}
\begin{equation}
\label{chandrasigma}
\text{log P = }
\begin{cases}
 0.1145\theta - 0.4957 \text{ log C} + 0.1932 & \\\text{if } 0.0000 \leq \text{log C} \leq 2.1393\\
0.0968\theta - 0.2064 \text{ log C} - 0.4260 & \\\text{if } 2.1393 \leq \text{log C} \leq 3.3\\
\end{cases}
\end{equation}
\end{center}

The corresponding 95 per cent X-ray error circle can vary from $\sim 0.8$ arcsec for small off-axis angles up to 19.6 arcsec for faint sources detected at a large off-axis angle. In addition to the X-ray uncertainties we also include a term for the mean uncertainty of the astrometric fit of the optical observations, and we add the total X-ray and optical terms in quadrature.

\begin{center}
\begin{figure}
  \includegraphics[width=7.5cm, keepaspectratio]{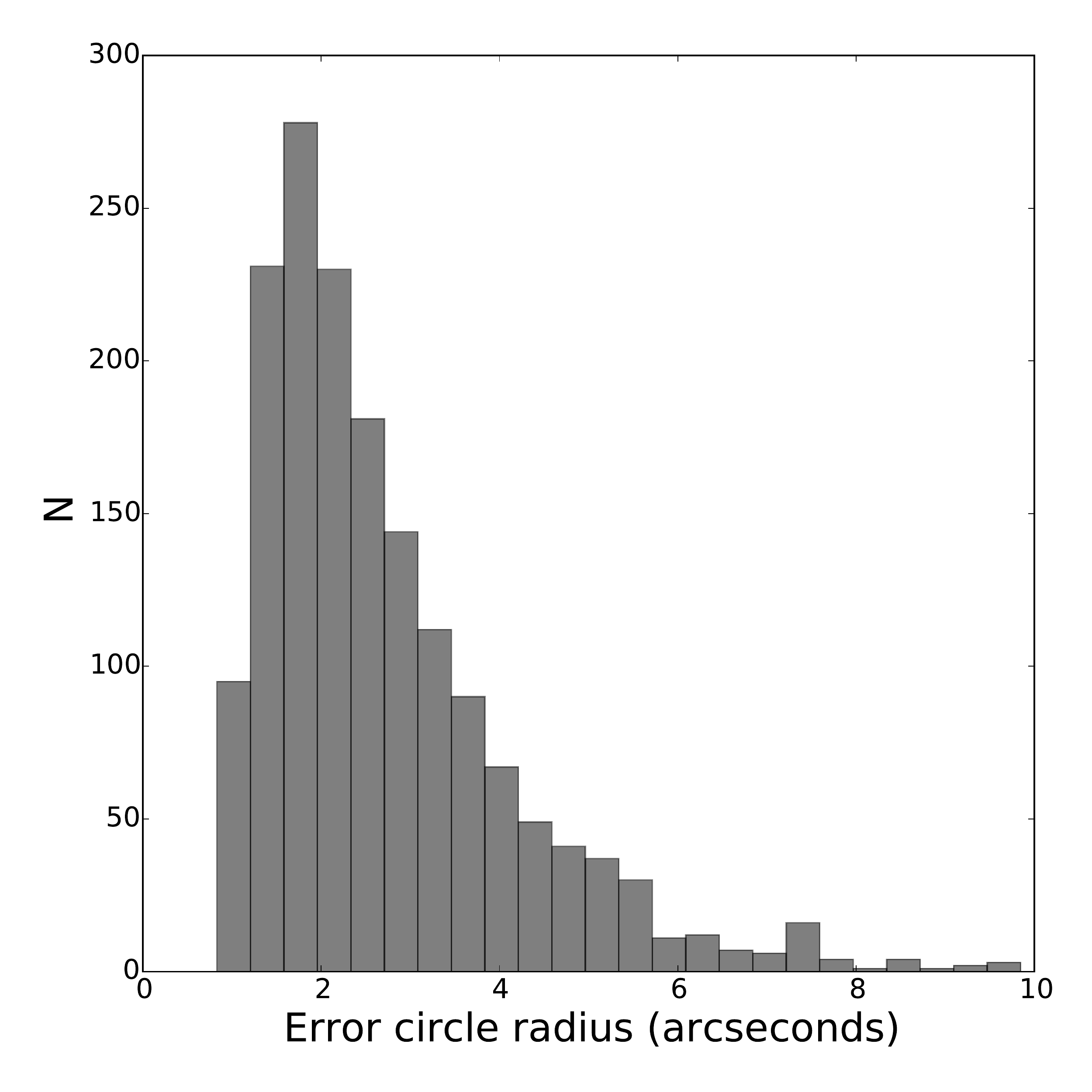}
  \caption{Distribution of the 95 per cent error circle radius of the X-ray positions. There are 6 extreme outliers, not shown here, for which the error circle is larger than 10 arcsec. The median 95 per cent error circle has a radius of R$_{95}$ = 2.33 arcsec. }
  \label{fig:r2sigma}
\end{figure}
\end{center}

In Figure \ref{fig:r2sigma} we show the distribution of 95 per cent error circle radii for the X-ray sources. There are 6 sources with uncertainties higher than 10 arcsec.
The median 95 per cent error circle has a radius of R$_{95}$ = 2.33 arcsec.

\subsection{False alarm probabilities}
\label{sec:fap}
\begin{center}
\begin{figure}
  \includegraphics[width=8.5cm, keepaspectratio]{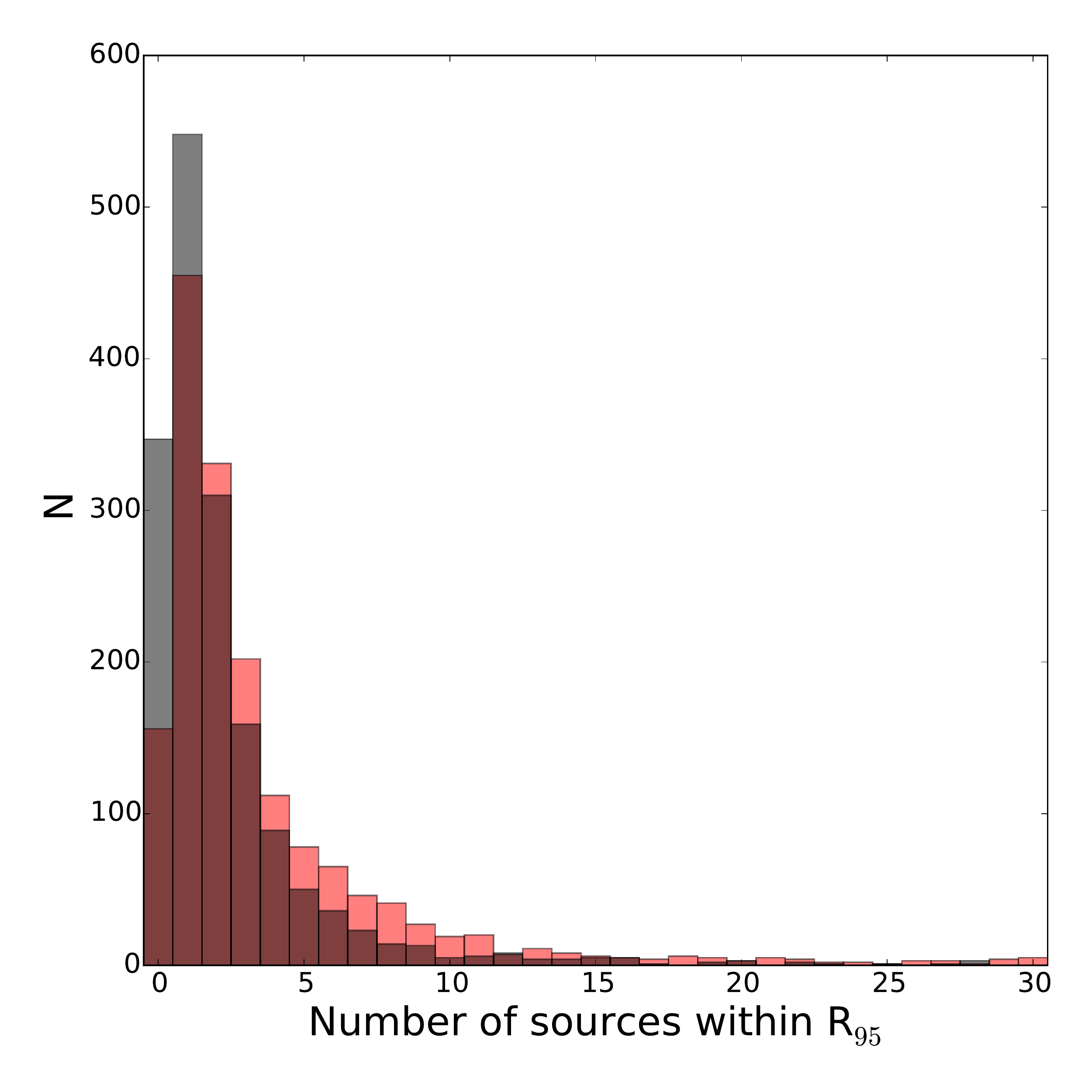}
  \caption{Distribution of the number of potential counterparts within R$_{95}$ for the most (grey) and least (red) conservative samples. The first bin represents sources without a candidate optical counterpart in the error circle.}
  \label{fig:nrofmatches}
\end{figure}
\end{center}

We take the X-ray sample and cross-match the positions with the optical catalogue, retaining all matches within the 95 per cent error circle. 
The 95 percentile is motivated by the fact that the relative number of additional true counterparts with respect to the number of false positives will increase with increasing distance towards the best-fit X-ray position (because the area of the enclosing circle increases).
In Fig. \ref{fig:nrofmatches} we plot the number of optical sources in the X-ray error circle for the most and least conservative optical samples in grey and red, respectively. If an X-ray source is present on multiple fields, we use the average number of unique optical sources within the error circle over all fields to avoid counting sources multiple times.
 
We find at least one potential optical counterpart within the combined astrometric errors for 954 (1160) out of 1640 X-ray sources using the most (least) conservative optical sample. 
The number of optical sources in the latter sample is a factor of 2 higher than in the most conservative one. We see that the bins with 0 or 1 potential counterpart get redistributed in a tail towards higher numbers of potential counterparts as we change from the most to the least conservative optical sample. The number of X-ray sources with one unique counterpart decreases by $\sim$ 25 per cent when allowing sources detected only in the i$^{\prime}$-band.

This implies that crowding plays a more important role in the i$^{\prime}$-band relative to the r$^{\prime}$-band. In general, the fact that multiple potential counterparts are present within the error circle for more than 75 per cent of the sources will have an important effect on our results, and crowding will lead us to find matches that are random alignments.

We now quantify the probability of an X-ray source being randomly aligned with an optical source of a given magnitude. These sources are contaminants in our samples, so it is of interest to estimate the fraction of interlopers as described below.
We use the stellar densities in the immediate environment of the X-ray position to evaluate the number of sources expected to be present in to the background. 
We estimate the optical stellar densities by binning the total number of stars in circular areas centered on the X-ray error circles in bins of 0.5 mag. We are interested in the local stellar density, hence we do not want to venture too far from the X-ray positions as this will increase the sensitivity to background variations of e.g. the reddening, which can severely affect the source densities. However, we want a robust estimate, so a small area is not desireable either. The radii of the circular areas are chosen such that the median number of empty magnitude bins does not exceed 25 per cent. They amount to radii of 100, 90 and 75 arcsec, where the largest radius belongs to the most conservative optical sample (which has the lowest stellar densities). We have shown (Sec. \ref{sec:comparisonsimobs}) that large reddening gradients exist on scales of $\sim$ 10 arcminutes or smaller, so these radii are small enough that our stellar densities will not be affected significantly by reddening variations between the X-ray source position and the location where we determine the optical source densities.

We compute the number of sources that we expect to fall in an area the size of our error circle assuming the computed stellar densities are uniform over the field. In that case, the expected number of background sources in each magnitude bin within R$_{95}$ is:
\begin{equation}
Y_m = \pi R_{95}^2 N_m
\end{equation}
where N$_m$ is the background stellar density per magnitude bin.
The false alarm probability (FAP) is quantified as the probability of finding one or more random sources in the error circle, given that we expect a certain number of sources (Y$_m$, estimated from the stellar density) to be present by chance. Assuming Poissonian statistics, we get:
\begin{equation}
\text{FAP} = 1 - \text{Pr(0, Y$_m$)} = 1 - e^{-\text{Y$_m$}}
\end{equation}

In many cases the cross-matching yields more than 1 potential counterpart within R$_{95}$. We perform the analysis described above using all possible counterparts, i.e. for all optical sources that fall within the error circle of an X-ray position, we calculate the FAP. If sources are present on multiple fields, we include all optical matches in our analysis. This means that sometimes the same unique optical source will have multiple FAP values depending on which field it was detected on\footnote{The FAP of a potential counterpart detected in multiple fields can change because the magnitude of the source is not necessarily identical if detected more than once.}. We select the most likely counterpart as the source which has the lowest FAP for each X-ray source. If the FAP value of a potential optical counterpart found on multiple fields is lowest for all cases, this provides confirmation that that source is the most likely counterpart.

\begin{figure}
  \includegraphics[width=\columnwidth, keepaspectratio]{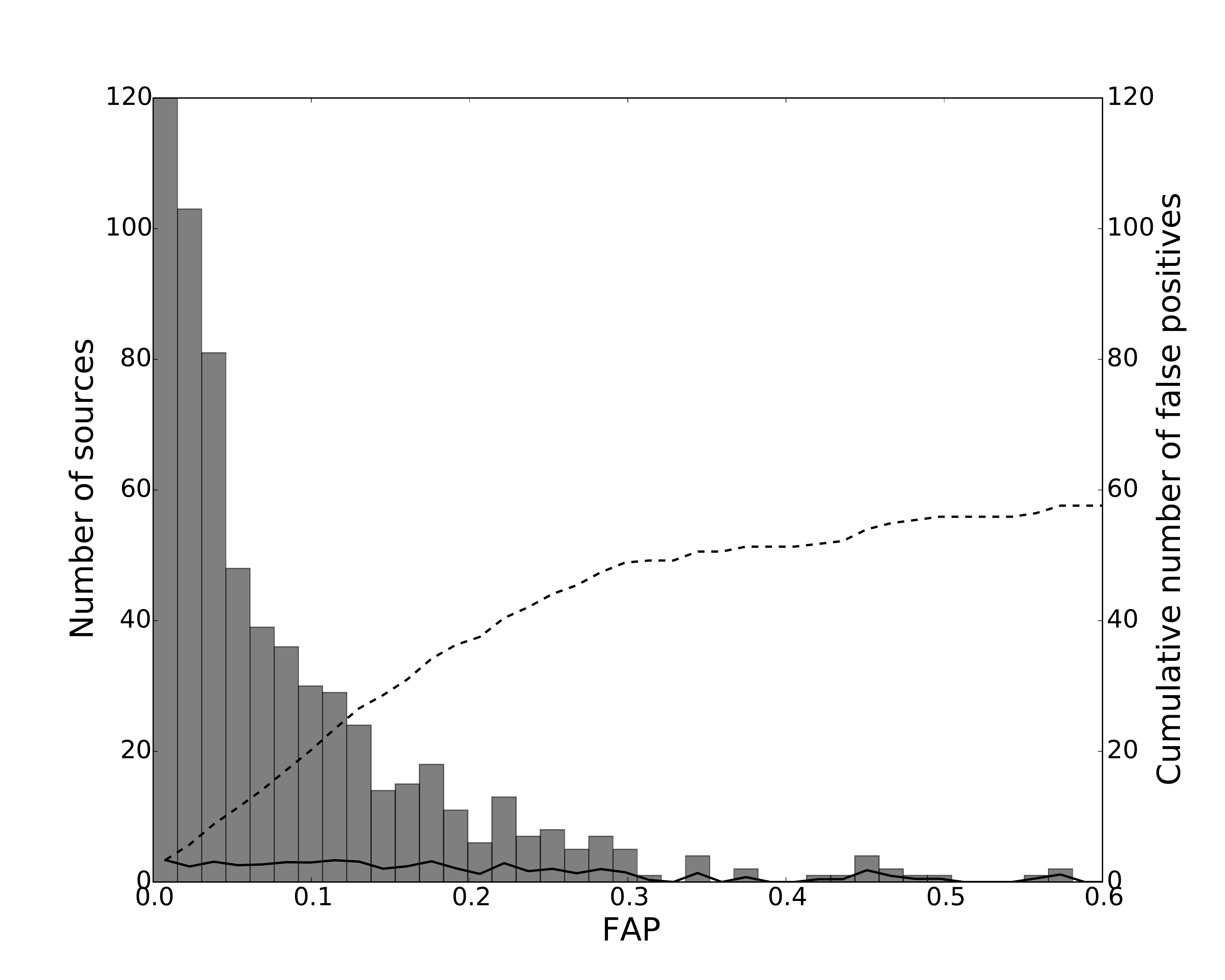}
  \caption{Histogram of FAP values of the most likely counterparts to the ensemble of X-ray sources when considering only sources that are detected in all three bands. The solid line represents the number of expected false positives in each bin, defined as the number of sources multiplied by the FAP value at the centre of the bin. The dashed line indicates the cumulative number of expected false positives up to a given FAP value.}
  \label{fig:Nfap}
\end{figure}

We use these most likely counterparts to estimate the number of sources that are expected to be chance alignments. Figure \ref{fig:Nfap} shows the distribution of FAP values of the most likely counterparts in the least conservative sample. We interpret the FAP of a given source as the chance that this most likely optical counterpart is present due to chance. For example, if we have 10 potential counterparts with FAP=0.1, we expect that 1 of those 10 sources is an interloper. Using the distribution of FAP values, we estimate the number of false positives per bin by multiplying the number of sources with the FAP value at the centre of each bin. This number is indicated for each bin by the solid line in Figure \ref{fig:Nfap}. The dashed line represents the cumulative number of expected false positives up to a given FAP. For the most and least conservative samples, we expect respectively 59 and 106 interlopers to be present, which amounts to 6 and 9 per cent of all X-ray sources for which we find a potential counterpart. 

In the above analysis, we have assumed that the field we use to determine the background stellar densities is representative for the stellar population in the X-ray error circle. However, in Section \ref{subsec:brightsources} we showed that the source detection algorithm is affected by the presence of bright stars. Because there is an overdensity of bright optical sources in the X-ray error circles, this may introduce a bias in our estimate for the contamination due to interlopers. We note here that if we remove all error circles containing stars brighter than i$^{\prime}\leq16$, the amount of expected false positives decreases from 106 to 95, hence the contribution of bright stars to the false positives is small and only influences our results at the per cent level.

\subsection{Target selection for spectroscopic follow-up}
\label{sec:likelihood}
Regarding the question of which optical source to target with spectroscopic observations for classification, we base ourselves on the method outlined by \citet{Sutherland1992}. These authors quantify a likelihood ratio for each optical source within the error circle of an X-ray source as the ratio of the probability of finding the X-ray source at a certain position within the error circle to the chance of finding a background optical source at the same location. This probability is primarily determined by the model for the PSF of the Chandra satellite.
For this particular method we will not restrict ourselves to the 95 per cent error circle, because we want to find a counterpart for as many X-ray sources as possible, and we do not want to miss a priori a number of counterparts to X-ray sources for which we statistically expect the X-ray source to fall outside R$_{95}$. We will therefore use a 4R$_{\sigma}$ error circle (99.96 per cent CL) for this analysis. The method that we employ ensures that sources at larger offsets from the nominal X-ray position will receive a lower likelihood ratio.

We consider an X-ray source with equal positional uncertainties in right ascension and declination, with Gaussian distributions g(x,y) in both directions:\\
\begin{equation}
\text{g(x,y)} = \frac{1}{2\pi\sigma^2} e^{-\frac{x^2+y^2}{2\sigma^2}}
\end{equation}
The probability density of counterparts at an offset ($\Delta$x, $\Delta$y) is g($\Delta$x, $\Delta$y), in units of mag$^{-1}$ arcsec$^{-1}$. We assume that the probability distribution of the true counterpart located at a distance r = d/$\sigma$ (where d$^2$ = ($\Delta$x)$^2$ + ($\Delta$y)$^2$) from the best-fit position follows a Rayleigh distribution:
\begin{equation}
p(r)dr = r e^{-\frac{r^2}{2}} dr
\end{equation}

The likelihood ratio for objects in the 4$\sigma$ error circle is the relative probability for a given candidate of finding the true counterpart at a certain offset and magnitude versus that of finding a background source at the same magnitude and offset:
\begin{equation}
\text{L(m,r)} =\frac{r e^{-\frac{r^2}{2}} dr}{2\pi r N_m dr} = \frac{e^{-\frac{r^2}{2}}}{2\pi N_m}
\end{equation}

In the above, we have ignored the factor Q defined in \citet{Sutherland1992} to account for the probability that the X-ray source actually has an optical counterpart within the survey detection limits, or for example the prior on the probability that the counterpart is in a certain magnitude bin, that it is located at a certain offset or has a specific colour index. It is in principle possible to extend this analysis to include the results of a first iteration to determine the priors on these probabilities (see for example \citet{Naylor2013} for an extensive discussion).
It would then be possible to quantify a lower cut-off for the likelihood ratio value for optical sources: when the chance of the source being an interloper is very high we could decide not to spend observing time on it. An extended analysis including priors would mainly impact the absolute likelihood ratio values of candidate counterparts, not the relative differences between sources within the same error circle. An extended analysis is outside the scope of this paper, so we set Q = 1 independent of the candidate counterpart magnitude and offset.
We then select the optical source with the highest likelihood ratio as the most likely counterpart and best target for spectroscopic follow-up.

\subsubsection{Detailed example of the likelihood ratio results}
\begin{figure*}
\minipage{0.5\textwidth}
  \includegraphics[width=\linewidth]{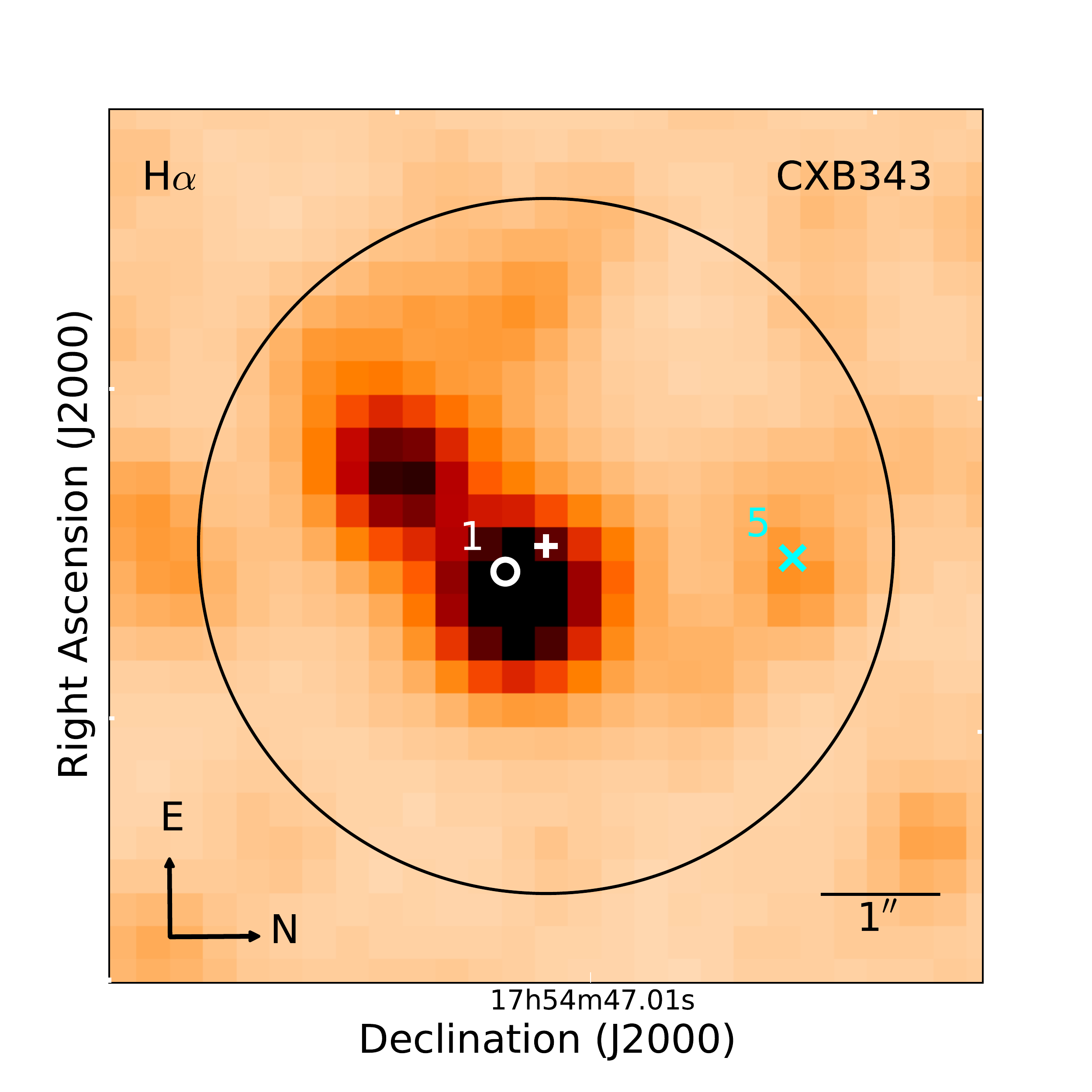}
\endminipage\hfill
\minipage{0.5\textwidth}
  \includegraphics[width=\linewidth]{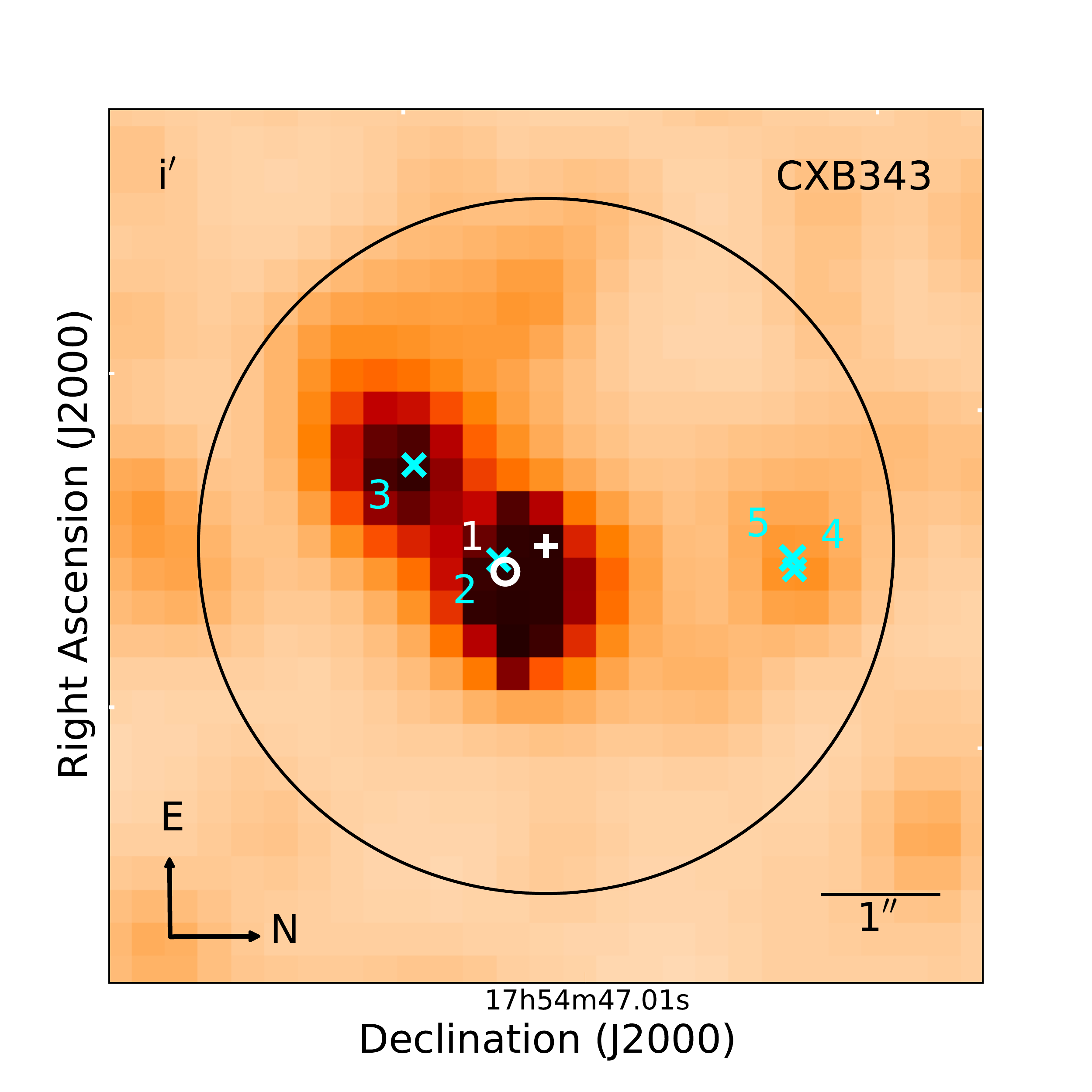}
\endminipage
\caption{Example of the region around an X-ray source (CXB343) for which we found multiple counterparts. The left panel shows the H$\alpha$ image, and overplotted the sources present in the most conservative optical sample, while the right panels shows the i$^{\prime}$-band image and all sources in the least conservative sample. The X-ray position is marked with a plus sign, and the 4R$_{\sigma}$ error circle (with a radius of 2.83 arcsec) is drawn in black. Cyan crosses mark the positions of the objects listed in Tab. \ref{tab:cxb343}. The white small circle is the source with the highest likelihood ratio, i.e. the most likely counterpart. If multiple crosses appear on the same source, this means that the source was detected on multiple detectors and/or overlapping fields. }
\label{fig:cxb343}
\end{figure*}
As an illustration, we show the results of the analysis described above in the case of CXB343. 
\begin{table*}
 \centering
  \caption{List of all potential optical counterparts within the error circle (with radius 2.83 arcsec) of CXB343. The source is located on field S10, and the X-ray coordinates are ($\alpha$, $\delta$) = (268.69700, -29.34595). The list is sorted from highest to lowest likelihood ratio (L). The offset d between the centre of the error circle and the optical source position, d, is given in units of R$_{\sigma}$. The rms of the astrometric solution is 0.11 arcsec for these frames.}
  \begin{tabular}{cccccccccccccccc}
  \hline
$\#$ & CCD&	RA ($^{\circ}$)	&	Dec ($^{\circ}$)& r$^{\prime}$	&	$\sigma_{r^{\prime}}$	&	i$^{\prime}$	&	$\sigma_{i^{\prime}}$&	H$\alpha$	&	$\sigma_{H\alpha}$		&d ($\sigma)$		& 	L		&	r$^{\prime}$ flag&	i$^{\prime}$ flag&	H$\alpha$ flag\\
\hline
1&6&268.696930&-29.346039&16.59&0.05&13.97&0.05&16.21&0.05&0.42&74.80&-1&-9&-1\\
2&7&268.696960&-29.346054&	16.59&0.05&15.94&0.05&16.08&0.05&0.50&48.78&-1&-9&1\\
3&7&268.697205&-29.346243&	0.00&0.00&16.27&0.05&	0.00&0.00&1.37&3.83&0&-1&0\\
4&7&268.696930&-29.345392&	19.49&0.08&18.23&0.08	&18.92&0.08&2.32&0.26&-1&-1&-1\\
5&6&268.696930&-29.345388&19.48&0.09&18.28&0.09&18.91&0.09&2.34&0.24&-1&-1&-3\\
  \hline
  \end{tabular}
  \label{tab:cxb343}
\end{table*}
In Figure \ref{fig:cxb343} we show the image stamp of this source. The left image is the H$\alpha$ observation, and overplotted are all the sources found within the error circle of the conservative optical sample. On the right we show the i$^{\prime}$-band image, and overlay the potential counterparts found in the least conservative sample. Table \ref{tab:cxb343} lists the properties of all the optical counterpart candidates found within the error circle, which is overplotted in black and has a 4$\sigma$ radius of 2.83 arcsec. The plus sign marks the best-fit X-ray position, and each cross marks a source that corresponds to an entry in Table \ref{tab:cxb343}. If 2 crosses are drawn on the same source, there are multiple detections. This can be in the offset exposure, on another CCD, or on an overlapping field (or a combination of these). The most likely counterpart (with the highest likelihood ratio) is denoted by a small white circle. In this case sources 1 and 2 in the table are the same object, seen on 2 different detectors. The analysis results in the highest likelihood ratio for this object in both observations, confirming that this is the most likely counterpart for the X-ray source. Looking at the magnitudes, the table reveals an apparent dimming of the counterpart by 2 magnitudes in the i$^{\prime}$-band, whereas there is no significant change in the other 2 bands. The flags show, however, that the source is saturated in both i$^{\prime}$-band observations, indicating that the apparent variability is likely caused by the unreliable photometry for saturated sources. \\
Source number 3, which is located in the south-east of the X-ray error circle, is nearly as bright as the most likely counterpart but it has a sharper PSF compared to other sources in the field and is flagged as noise-like in r$^{\prime}$ and H$\alpha$.
The other sources within the error circle are less feasible candidates as they are at larger distances and fainter magnitudes (increasing the likelihood of chance alignment) than sources 1 through 3, which is reflected in lower likelihood ratios.

\section{Results and discussion}
\label{sec:results}
\subsection{Comparison between FAP and likelihood ratio}
It is instructive to compare the results of the two methods we have described in the previous section to find the most likely optical counterparts to X-ray sources. Because the FAP method uses a smaller error circle, we cross-match these sources with the sources that have the highest likelihood ratio for their respective X-ray sources. We use a cross-matching radius of 0.5 arcsec, as we did in our optical samples. 

In agreement with our previous estimation of the contamination in the counterpart samples, we find that both methods yield the same optical source as the most likely candidate in 91 (88) per cent of the most (least) conservative counterpart candidates. The bulk of optical sources for which we find a different counterpart can be intuitively explained by recalling that the likelihood ratio takes into account the distance from the centre of the error circle, while the FAP values do not. 
These sources constitute error circles that contain a bright optical counterpart candidate on the outskirts of R$_{95}$. Given that the density of bright stars is low compared to fainter objects, these sources naturally have lower false alarm probabilities. However, when we take into account the distance to the source in the calculation of the likelihood ratio, faint sources close to the centre of the error circle can obtain higher likelihood ratio values. 

We conclude that by comparison of the most likely counterparts based on the two different metrics we have adopted, we can expect about 10 per cent of sources to be chance alignments.

\subsection{Comparison of the counterpart populations}
\label{sec:ctparts}
\begin{table*}
 \centering
  \caption{First five entries of the most important information of the most likely optical counterparts. The radius of the error circle, R$_{4\sigma}$, is given in arcseconds. The offset d is given in units of R$_{\sigma}$. The number of unique optical sources within each error circle is also given. If the X-ray source is present on multiple fields, we quote the average of all fields (which explains the presence of non-integer numbers in this column). The final three columns denote the classification flag in each filter. The full tables for all three optical samples are available in the online material.}
  \begin{tabular}{cccccccccccccccc}
  \hline
ID & RA$_{opt}$ & Dec$_{opt}$ & r$^{\prime}$ & $\sigma_{r^{\prime}}$ & i$^{\prime}$ & $\sigma_{i^{\prime}}$ & H$\alpha$ & $\sigma_{H\alpha}$ & R$_{4\sigma}$ & d ($\sigma$) & L & \# & r$^{\prime}$ & i$^{\prime}$ & H$\alpha$\\\hline
CX2 & 264.368256 & -29.133936 & 18.37 & 0.02 & 16.78 & 0.01 & 17.96 & 0.02 & 1.55 & 0.60 & 65.90 & 1.50 & -1 & -1 & -1 \\
CX3 & 265.178589 & -28.302095 & 0.0 & 0.0 & 20.94 & 0.14 & 0.0 & 0.0 & 1.50 & 2.08 & 5.26 & 1.00 & 0 & -1 & 0 \\
CX5 & 265.038086 & -28.790514 & 19.13 & 0.02 & 17.94 & 0.01 & 18.09 & 0.02 & 1.47 & 0.63 & 85.37 & 1.00 & -1 & -1 & -1 \\
CX11 & 265.464233 & -27.039984 & 21.1 & 0.04 & 18.91 & 0.02 & 20.52 & 0.07 & 3.94 & 1.67 & 4.93 & 1.00 & -1 & -1 & -3 \\
CX15 & 266.692719 & -25.871557 & 19.32 & 0.02 & 18.17 & 0.02 & 18.85 & 0.02 & 2.58 & 1.65 & 4.74 & 1.00 & -1 & -1 & -1 \\
  \hline
  \end{tabular}
  \label{tab:xmatchresult}
\end{table*}
We now turn to exploring the properties of the most likely optical counterparts, i.e. those that have the highest likelihood ratio calculated using the method above. In total, we find counterparts to 1287 and 1480 X-ray sources for the most and least conservative optical samples, respectively. 

Table \ref{tab:xmatchresult} contains an example of the most relevant properties of the most likely counterparts. The full tables are available in the online material.

In Section \ref{sec:xmatch} we defined three different selection procedures for the optical samples used in our analysis, starting with a very conservative case with a stellar PSF detection in all bands. Subsequently we allow sources with only r$^{\prime}$ and i$^{\prime}$ detections and sources only detected in the i$^{\prime}$-band.

\begin{center}
\begin{figure}
  \includegraphics[width=8.5cm, keepaspectratio]{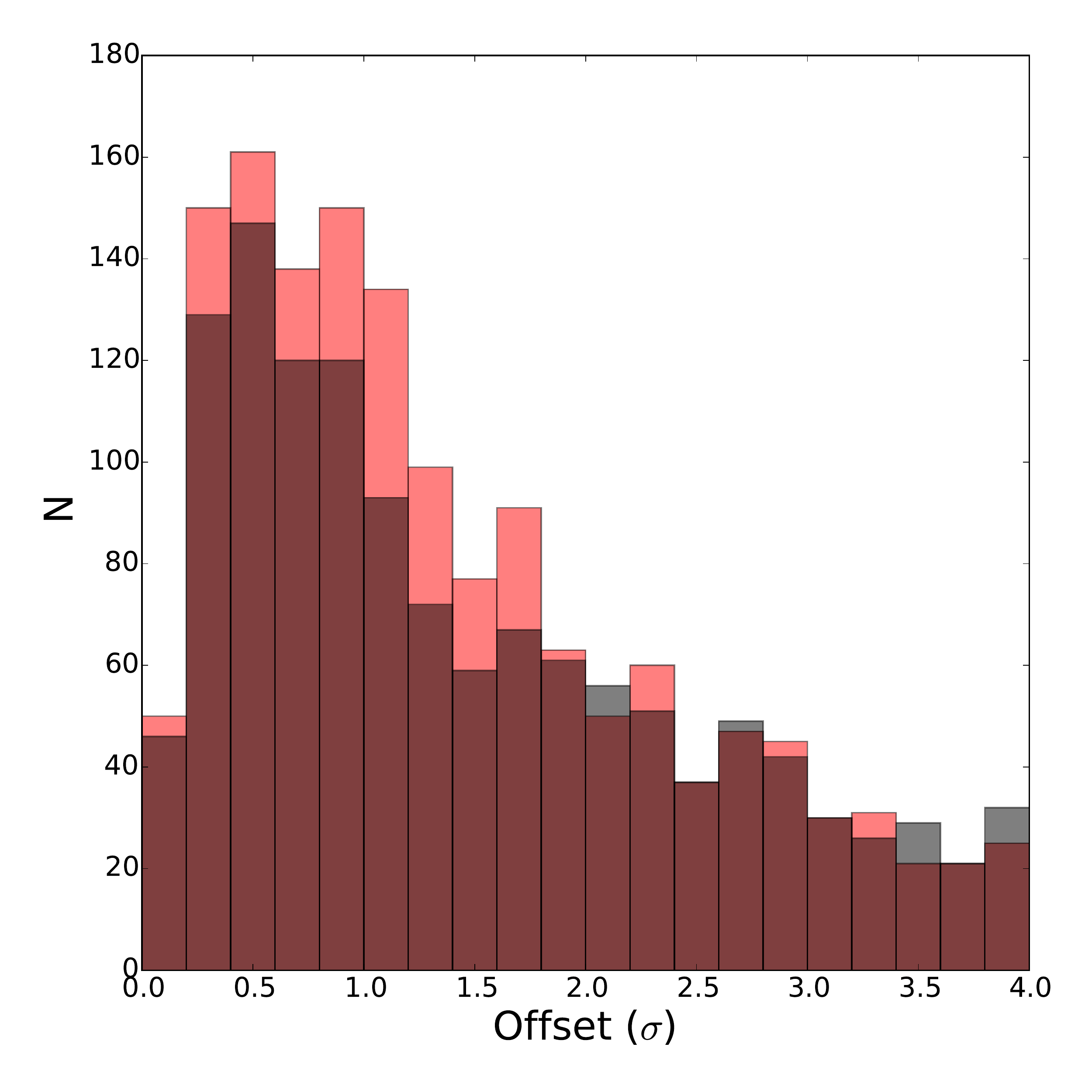}
  \caption{Distribution of the distances between the most likely counterparts and the best-fit X-ray position for the most (grey) and least (red) conservative samples. We normalised the offset to the 1$\sigma$ combined astrometric errors per source. 75 per cent of the most likely counterparts that are only detected in the i$^{\prime}$-band are found within 2$\sigma$ of the best-fit X-ray position.}
  \label{fig:dist2x}
\end{figure}
\end{center}

When we compare the results for the most likely counterparts, there are no large differences in their properties between the three samples (Table \ref{tab:resultmatch}). What one can expect is that when we include more sources in our respective optical samples, we find more candidate counterparts to X-ray sources. Because the observations in H$\alpha$ are shallower than in r$^{\prime}$ and i$^{\prime}$, we expect the median magnitude of the counterparts to become fainter. Due to the difference in reddening between the r$^{\prime}$ and i$^{\prime}$ bands, we expect that a significant fraction of counterparts will only have an i$^{\prime}$-band detection (specifically, those located more than a few kpc from Earth) because interstellar reddening is more important in the r$^{\prime}$-band. We see that indeed the median magnitude shifts towards fainter values while the offset between the most likely optical counterpart and the centre of the X-ray error circle becomes smaller. In Figure \ref{fig:dist2x} we show the distribution of offsets between the most likely optical counterpart and the best-fit X-ray position. 75 per cent of the most likely counterparts that are detected only in the i$^{\prime}$-band are found within 2$\sigma$ of the best-fit X-ray position.

\begin{figure} 
  \includegraphics[height=8.5cm, keepaspectratio]{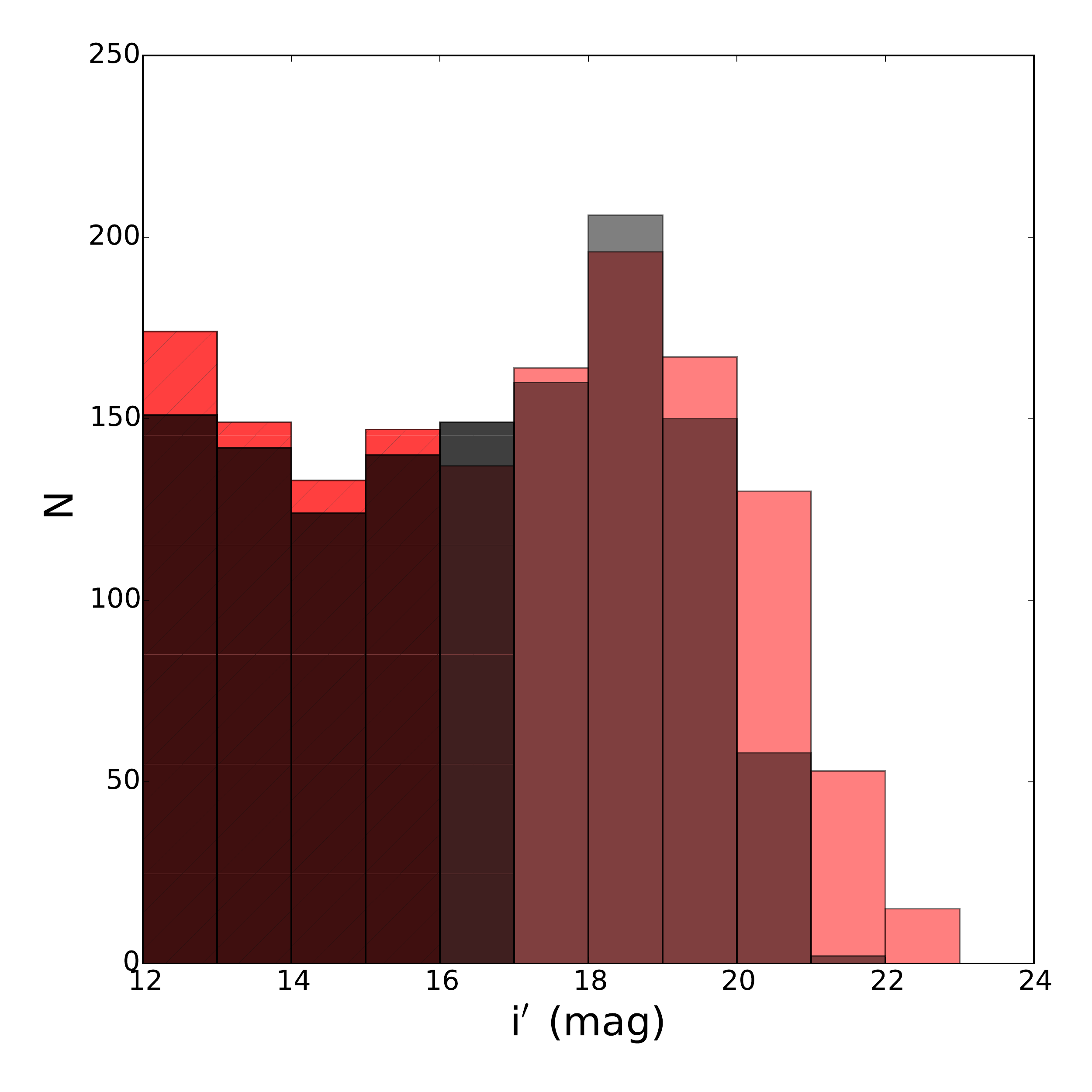}
  \caption{Distribution of the i$^{\prime}$-band magnitudes of the most likely counterparts in bins of 1 mag for the most (grey) and least (red) conservative samples. The distribution is dominated by a population of bright foreground sources that are flagged as saturated ($i^{\prime} \leq 16$, hatched). Additional candidate counterparts found in the least conservative sample are mostly faint ($i^{\prime} \geq 19$) sources. }
  \label{fig:magdistctpart}
\end{figure}

The magnitude distribution of the final sample of candidate counterparts is shown in Figure \ref{fig:magdistctpart} for the most (grey) and least (red) conservative cases.
When we look at the bright end of the magnitude distribution, there are a total of 754 candidate counterparts brighter than 17th magnitude in the $i^{\prime}$-band, or 51 per cent of all candidates. Of these 754, 584 sources have photometry that is flagged as saturated in $r^{\prime}$ and/or $i^{\prime}$.
As an order of magnitude estimation, we note that a K2V dwarf (M$_{r^{\prime}}$ = 6.3, see \citet{Allen2000} and the colour transformation taken from \citeauthor{Jester2005} \citeyear{Jester2005}) seen at $r^{\prime}$ = 17 mag is consistent with a source distance of $\sim$ 900 pc assuming A$_{r^{\prime}}$ = 1. This indicates that these sources comprise a population of foreground sources for which the extinction is low, and they are likely the real counterparts to the X-ray sources.

\begin{table}
 \begin{minipage}{\columnwidth}
 \begin{center}
  \caption{Comparison of the populations of the most likely counterparts using the different optical samples. We have excluded saturated sources. Offset denotes the offset between the centre of the X-ray error circle and the most likely optical counterpart in units of $\sigma$, the radius of the error circle. We quote the median magnitude and offset of the respective optical samples.}
  \begin{tabular}{lccc}
\hline
Sample & Conservative & no H$\alpha$ & only i$^{\prime}$ \\\hline
i$^{\prime}$ (mag) & 18.17 & 18.23 & 18.51 \\ 
Distance (R$_{\sigma}$) & 1.59 & 1.47 & 1.33  \\
Number & 763 & 814 & 901\\
  \hline
  \end{tabular}
  \label{tab:resultmatch}
  \end{center}
 \end{minipage}
\end{table}

Another indication that a large number of the most likely counterparts are foreground stars can be obtained by cross-matching the most likely counterparts of the least and most conservative samples within the astrometric errors and then identify sources for which the optical source with the highest likelihood ratio has changed between samples. We find that 447, or 30 per cent, of the sources have a different candidate counterpart. This means that as much as $\sim$ 70 per cent of the sources is the same for both samples, implying that they are not heavily affected by interstellar reddening since they are detected in all three bands and thus are likely located nearby.

\begin{figure}
  \includegraphics[width=\columnwidth, keepaspectratio]{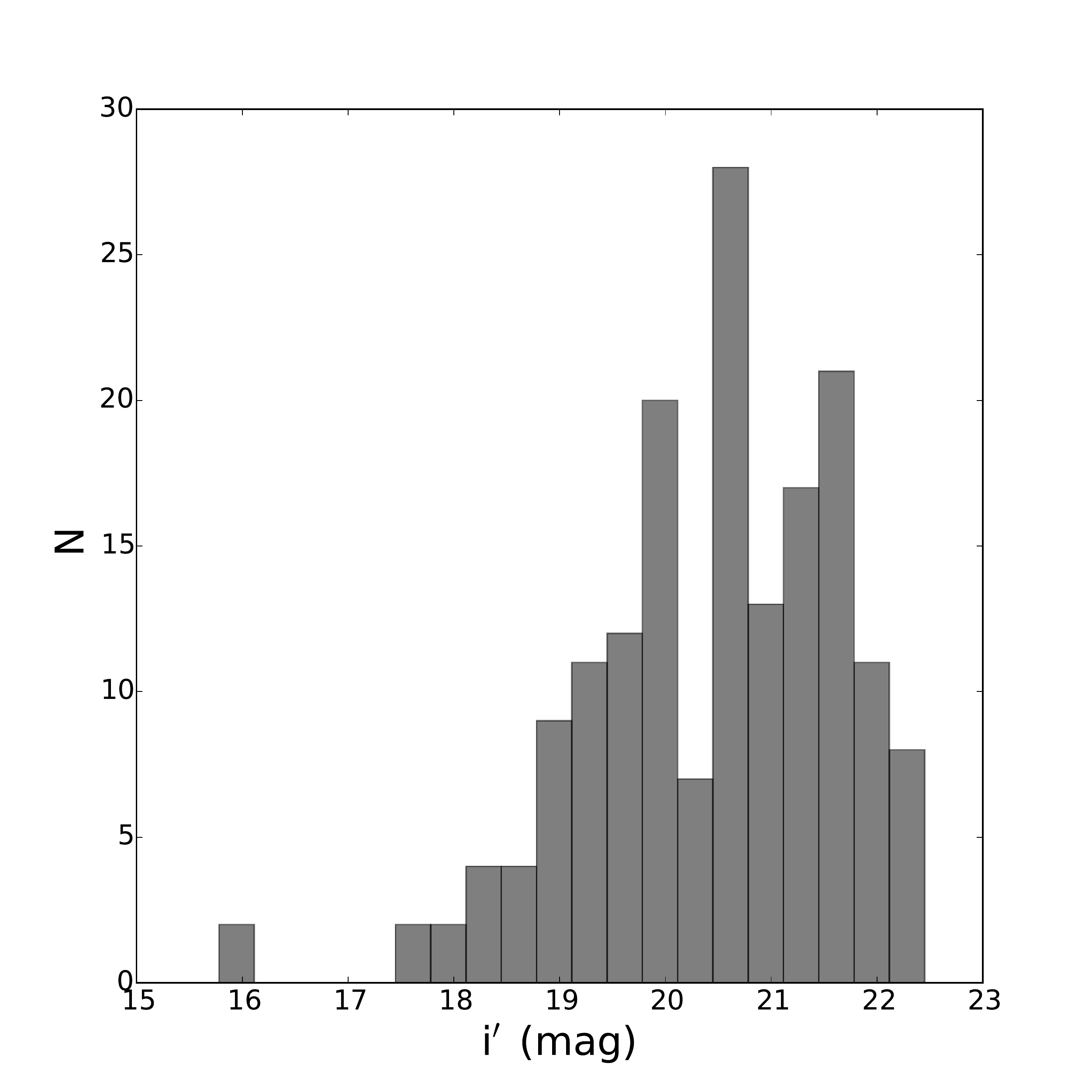}
  \caption{Magnitude distribution of additional candidate counterparts, found when comparing the least and most conservative samples within the optical astrometric errors. We have omitted sources flagged as saturated in any band, and only plot sources that are detected in the i$^{\prime}$-band, but not in the r$^{\prime}$-band. These 171 objects are excellent candidates for spectroscopic follow-up.}
  \label{fig:newcp_i_imag}
\end{figure}
In Figure \ref{fig:newcp_i_imag} we show the population of sources that are detected in the i$^{\prime}$-band, but have a non-detection in the r$^{\prime}$-band. The median i$^{\prime}$-band magnitude of these sources is 20.58, fainter than the median magnitude of the candidate counterparts in each of the optical samples. This population comprises 171 sources, and these objects are prime targets for spectroscopic follow-up. 
If we combine the median i$^{\prime}$-band magnitude with the median 5$\sigma$ detection limit in the r$^{\prime}$-band, we obtain a lower limit on the typical observed colour of r$^{\prime}-i^{\prime}=1.9$ for these sources. Assuming a system containing a K2V star, which has an intrinsic colour of r$^{\prime}-i^{\prime}=0.26$ (see the assumptions in Sec. \ref{sec:ctparts}), this implies a differential reddening in the r$^{\prime}$-band with respect to i$^{\prime}$ due to dust extinction of 1.65 mag. Following \citet{Schlegel1998}, we find that A$_i$=0.76A$_r$, so to reach a differential dust extinction of 1.65 mag the source needs to be reddened by $A_r=6.9$ mag. Using the reddening values from \citet{Gonzalez2012}\footnote{http://mill.astro.puc.cl/BEAM/calculator.php}, we estimate that the average extinction in the GBS fields towards the Bulge is $\sim$8 mag in the r$^{\prime}$-band. We conclude that counterpart candidates detected only in the i$^{\prime}$-band are likely located near the Galactic Bulge, hence X-ray bright, making them good CV and LMXB candidates. 

To illustrate this claim, we look in some more detail at CX3, the third brightest X-ray source (1850 X-ray photons detected) in the GBS catalogue. The position of CX3 coincides with that of SwiftJ1734.5-3027, a known NS LMXB showing long type I X-ray bursts \citep{Bozzo2015}. The source was detected as a hard X-ray transient by Swift in 2013, and \citet{Bozzo2015} place it at a distance of 7.2 kpc. We find only one optical source within the combined astrometric uncertainties, and identify it as the optical counterpart of the system. We detect the optical source with i$^{\prime}$=20.9, and have non-detections in r$^{\prime}$ (lower limit of r$^{\prime}$=22.5) and H$\alpha$. This yields an X-ray to optical flux ratio of $\frac{F_X}{F_{opt}} \sim 500$, where F$_{opt}$ was determined from the i$^{\prime}$-band magnitude. Such a large value is typical for X-ray binaries. The detection of this source in our (soft) X-ray and optical observations predates the discovery of the type I X-ray bursts by Swift by $\sim$ 7 years.

In addition to active (binary) stars, CVs and LMXBs \citep{Jonker2011}, we expect on the order of a few hundred background AGN in the X-ray sample (see the discussion in \citeauthor{Britt2014} \citeyear{Britt2014}). Most X-ray selected AGN are hosted by spheroids and bulge-dominated galaxies \citep{Povic2012}. Because they suffer from dust extinction through the Galactic Bulge, we expect that they will appear as point-like sources, and we are not biased against their detection as counterparts to X-ray sources. Some of these background AGN could be identified by their high X-ray to optical flux ratio, as extinction tends to increase this ratio for background sources. Analysis of the X-ray to optical flux ratios together with spectroscopic confirmation is required to identify the background AGN in our sample of counterpart candidates.

\section{Summary}
\label{sec:summary}
We present a deep optical catalogue of the Chandra Galactic Bulge Survey fields (GBS; \citeauthor{Jonker2011} \citeyear{Jonker2011}, \citeyear{Jonker2014}) in 3 filters (r$^{\prime}$, i$^{\prime}$ and H$\alpha$) consisting of at least 2 (offset) epochs per pointing. The catalogue contains $\sim$ 22.5 million unique objects, and more than 54 million source detections. The average 5$\sigma$ depth of the observations is 22.5, 21.1 mag in r$^{\prime}$ and i$^{\prime}$, respectively. The catalogue is complete down to r$^{\prime}$ = 20.2 and i$^{\prime}$ = 19.2. We determine the astrometric solutions in each frame to an rms of 0.15 arcsec or better, with an average astrometric rms of 0.04 arcsec. In addition to the source positions, magnitudes and their uncertainties, extensive auxiliary information is made available in the published catalogues. For example, for each observation we supply a classification based on the shape of the PSF in comparison with the global PSF properties of the sources in an image frame. We compare the observations to synthetic photometry, and validate that they are in qualitative agreement using colour-colour and colour-magnitude diagrams. All the data, including the processed images, single-filter catalogues and the merged catalogues are available through the Vizier database (http://vizier.u-strasbg.fr).

Using 3 different subsamples of this catalogue, we search for the optical counterparts to X-ray sources discovered in the GBS. The optical samples consist of i) a very conservative sample, containing objects with a point-like PSF in all bands, ii) a sample where we no longer require a detection in the H$\alpha$ observations (because these are shallower than the broad-band observations), and iii) a sample where we require only an i$^{\prime}$-band detection. The last sample is motivated by the fact that the dust extinction along the observed lines of sight can be very high, hence the counterparts will be highly reddened if the X-ray source is located in the Bulge. Because each of these samples contains more than 8 million sources, it is not trivial to identify the most likely counterpart for each X-ray source. 

We compare the optical observations in regions around the X-ray sources with (pseudo-)random regions in the sky (to minimise the effect of interstellar extinction variations on the observed properties), and find that there are more bright stars in the vicinity of X-ray sources, indicating that the bright counterparts we find are likely to be the real. We also find that the optical sources appear to be clustered close to the X-ray positions with respect to random positions on the sky. This suggests that the candidate counterparts we find are unlikely to be random matches due to chance alignments (this is true for all candidates, regardless of their brightness). Using a false alarm probability analysis, we estimate the contamination of optical counterparts due to chance alignments based on the local stellar densities. We expect that $\sim$ 10 per cent of the candidate counterparts are interlopers. 

Because the optical colour information alone is not sufficient to unambiguously classify the X-ray sources, we determine which counterpart is the best candidate for spectroscopic follow-up using the likelihood ratio technique. This takes into account the distance from the centre of the X-ray error circle and the local (optical) stellar densities. 
For the whole sample of X-ray sources, we find 1287, 1345 and 1480 counterparts for the three respective samples. 754 sources have $\text{i}^{\prime} \leq $ 17 (of which 584 are saturated in our observations), indicating that they constitute a population of bright foreground sources which are probably the real counterparts. Comparing the most likely counterparts between the most and least conservative optical samples, we find that 447 sources have a different counterpart between samples. 171 sources are detected in the $\text{i}^{\prime}$-band but not in the r$^{\prime}$-band, and have magnitudes that are significantly fainter than the global population of counterpart candidates. This indicates that they comprise a separate population from the foreground sources. This also implies that, as expected \citep{Jonker2011}, a significant number of counterparts ($\sim$12 per cent) are detected only in the i$^{\prime}$-band. These counterpart candidates are either intrinsically red and faint, or the X-ray sources are located at large distances and suffer from interstellar extinction. 

Spectroscopic and photometric follow-up for these objects is either planned or ongoing, and this work will serve as the basis for future follow-up observations to constrain the nature of the identified systems. In a future article we will use these data to identify all H$\alpha$ emission and absorption line candidates, and perform spectroscopic follow-up of selected optical counterparts.

\section*{Acknowledgements}
We would like to thank the referee for his/her comments which helped to improve the manuscript. We thank professors P. Groot and F. Verbunt for a useful discussion on the Chandra astrometric uncertainties and false alarm probabilities. PGJ acknowledges support from European Research Council Consolidator Grant 647208. Based in part on observations at Cerro Tololo Inter-American Observatory, National Optical Astronomy Observatory (2006A-0086, PI: Jonker), which is operated by the Association of Universities for Research in Astronomy (AURA) under a cooperative agreement with the National Science Foundation. 


\bibliographystyle{mn2e.bst}
\bibliography{bibliography}

\label{lastpage}

\end{document}